# Two-Way Partial AUC and Its Properties

Hanfang Yang* Kun Lu† Xiang Lyu‡ Feifang Hu§

**Abstract**

Simultaneous control on true positive rate (TPR) and false positive rate (FPR) is of significant importance in the performance evaluation of diagnostic tests. Most of the established literature utilizes partial area under the receiver operating characteristic (ROC) curve with restrictions only on FPR, called FPR pAUC, as a performance measure. However, its indirect control on TPR is conceptually and practically misleading. In this paper, a novel and intuitive performance measure, named as two-way pAUC, is proposed, which directly quantifies partial area under the ROC curve with explicit restrictions on both TPR and FPR. To estimate two-way pAUC, we devise a nonparametric estimator. Based on the estimator, a bootstrap-assisted testing method for two-way pAUC comparison is established. Moreover, to evaluate possible covariate effects on two-way pAUC, a regression analysis framework is constructed. Asymptotic normalities of the methods are provided. Advantages of the proposed methods are illustrated by simulation and Wisconsin Breast Cancer Data. We encode the methods as a publicly available R package **tpAUC**.

**Keyword**: ROC curve; True positive rate; False positive rate; Discrimination capability; Diagnostic test.

*Corresponding Author. Associate Professor, School of Statistics, Renmin University of China, China. E-mail: hyang@ruc.edu.cn.
†PhD Student, Department of Operations Research and Financial Engineering, Princeton University, Princeton, NJ, USA.
‡PhD Student, Department of Statistics, Purdue University, West Lafayette, IN, USA.
§Professor, Department of Statistics, George Washington University, Washington, D.C., USA.

# 1 Introduction

In clinical practice, it is crucial to separate diseased subjects from non-diseased ones by conducting diagnostic tests. Among all performance measures for such tests, ROC curve is the most widely used one, which plots FPR versus TPR over all possible threshold level (1; 2). Area Under the ROC Curve (AUC) summarizes the performance information in the curve across all thresholds. It is equivalent to the probability of a randomly selected diseased object being ranked higher than a randomly selected non-diseased object by a classifier (3). In practice, diagnostic tests with high FPR result in significant economical expense: a great proportion of non-diseased candidates would exhaust the scarce resource of medical therapies. In addition, when diagnosing a lethal disease, failing to correctly identify severe diseased subjects (low TPR) will cause serious ethical consequences. As a result, FPR and TPR need to be simultaneously maintained at low and high levels respectively, so that uneconomical and unethical regions are ruled out from AUC. Most of the established literature of the partial area under ROC curve (pAUC) adopts indirect methods to assess the region of interest with TPR constraint. In particular, one of the most popular methods is to set lower and upper restrictions on FPR, which is named FPR pAUC and defined as FPR pAUC$(p_1, p_2) := \int_{p_1}^{p_2} ROC(t) dt$ and $p_1$ ($p_2$) is an lower (upper) constraint on FPR. See (4; 5; 6; 7; 8; 9; 10). The lower bound of FPR, $p_1$, is to indirectly maintain the lower bound of TPR at $ROC(p_1)$. Meanwhile, $p_2$ is to restrict FPR from being too high. However, regarding area under ROC in economical and ethical region, such a definition incorporates the redundant area below TPR constraint (see Figure 1 and Remark 3.1). This leads FPR pAUC to suffer inefficiency and inaccuracy in performance evaluation. Therefore, the need for a direct and practical performance measure of the region with high TPR and low FPR arises in clinical research.

In this paper, we design a novel performance measure, named as two-way partial AUC (two-way pAUC), which is a flexible tool to control explicit TPR and FPR restrictions. Unlike utilizing an artificial FPR lower bound to indirectly control acceptable TPR, two-way pAUC provides a straightforward measure to substitute existing methods with independent vertical and horizontal limits. Due to this natural principle, two-way pAUC is convenient and intuitive for implementation and interpretation (refer to Figure 1). In addition, as shown in Section 2, two-way pAUC enjoys efficiency and accuracy against FPR pAUC. To estimate two-way pAUC, we propose a nonparametric estimator. Asymptotic normality of the estimator is derived to construct confidence intervals. In classifier comparison, the difference of performance measure is a popular criterion to select the dominated classifier. Built on the estimator, we establish a bootstrap-assisted method to test the difference of two correlated two-way pAUCs. Furthermore, the performance of a classifier is usually affected by underlying factors in clinical practice (3; 6). To evaluate their effects on two-way pAUC, we propose a regression framework constructed by generalized linear model on conditional two-way pAUC with the covariates. Asymptotic justifications of regression parameters are established.

The rest of paper is organized as follows. Graphical representation is demonstrated in Section 2. In Section 3, we define two-way pAUC and propose its nonparametric estimator and regression framework. Theoretical properties are provided in Section 4. Simulation study and real data analysis are conducted in Section 5 and 6. Concluding remarks and further discussion are presented in Section 7.

## 2 Graphical representation

Graphical representation of two-way partial AUC is the shaded region (area A) in Figure 1. Note that the shaded region is directly determined by explicit TPR ($\geq q_0$) and FPR ($\leq p_0$) restrictions. In contrast, FPR pAUC (area A+B) indirectly maintains acceptable TPR by setting a FPR lower bound (green dotted line in Figure 1; see Remark 3.1 for its mathematical description), thus incorporates the redundant area $B$ below TPR constraint $q_0$. This violates the original intention of preventing TPR from low region. The following two scenarios illustrate incorrect decision-making. Note that each ROC curve corresponds to a classifier.

**Scenario 1:** In Figure 2a, $ROC_1$ dominates $ROC_2$ over every aspect of FPR and TPR. With economical and ethical concerns, only pAUC where TPR is greater than 0.5 and FPR is less than 0.5 is of interest. Two measures, two-way pAUC and FPR pAUC, are utilized to assess performance within the given region. Note that the lower bounds on FPR in FPR pAUC at two different ROC curves are respectively determined by the pre-specific TPR lower bound. Hence, as shown in Figure 2a, FPR pAUCs of $ROC_1$ and $ROC_2$ are the areas of $S_3 + S_4$ and $S_1 + S_2 + S_3 + S_4$ respectively, and corresponding two-way pAUCs are $S_3$ and $S_1 + S_3$. The difference of two classifiers' discrimination capabilities is $S_1$ by two-way pAUC. In contrast, the difference is $S_1 + S_2$ by FPR pAUC. Due to that $S_2$ is below TPR constraint, it should not be taken into consideration. The redundant unethical area, $S_2$, distorts the comparison of two classifiers. Thus, two-way pAUC with direct restrictions is more efficient (links to statistical power) than FPR pAUC (refer to numerical study in Section 5).

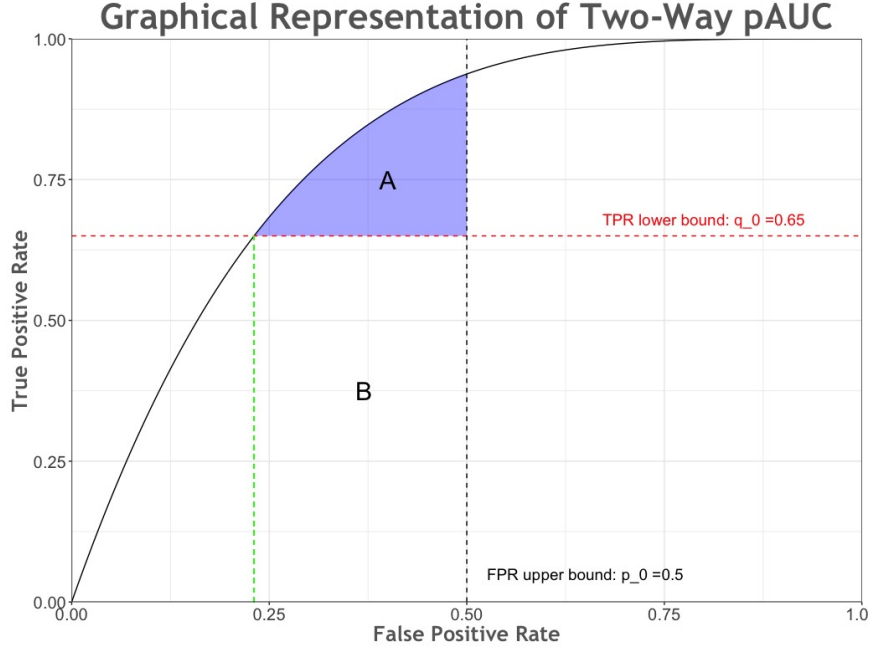

Figure 1: Two-way pAUC denotes the area of shaded region A. This shaded region is directly determined by explicit FPR upper bound $p_0$ ($= 0.5$) and TPR lower bound $q_0$ ($= 0.65$). In contrast, FPR pAUC denotes the area of both region A and B. Its indirect FPR lower bound (green dotted line) is determined by the TPR lower bound $q_0$.

**Scenario 2:** In Figure 2b, the region is of interest where TPR is larger than 0.6 and FPR is smaller than 0.6. $ROC_1$ has better discrimination capability in the region than $ROC_2$ (due to $S_6 > S_5$, as shown in Figure 2b). Two-way pAUCs for $ROC_1$ and $ROC_2$ equal $S_6 + S_7$ and $S_5 + S_7$, and FPR pAUCs for $ROC_1$ and $ROC_2$ are $S_6 + S_7 + S_9$ and $S_5 + S_7 + S_8 + S_9$, respectively. The difference of two-way pAUCs ($S_6 - S_5 > 0$) suggests that $ROC_1$ has better discrimination capability in the restricted region. However, when using FPR pAUC, a weighty and redundant area $S_8 (> S_6)$ is considered, which leads $ROC_2$ to be selected (due to $S_6 - S_5 - S_8 < 0$). Since the redundant region (below TPR restriction), which is incorporated by FPR pAUC, could distort the comparison result of discrimination capabilities to the opposite conclusion, two-way pAUC is more accurate. Real data application (refer to Section 6) supports the argument.

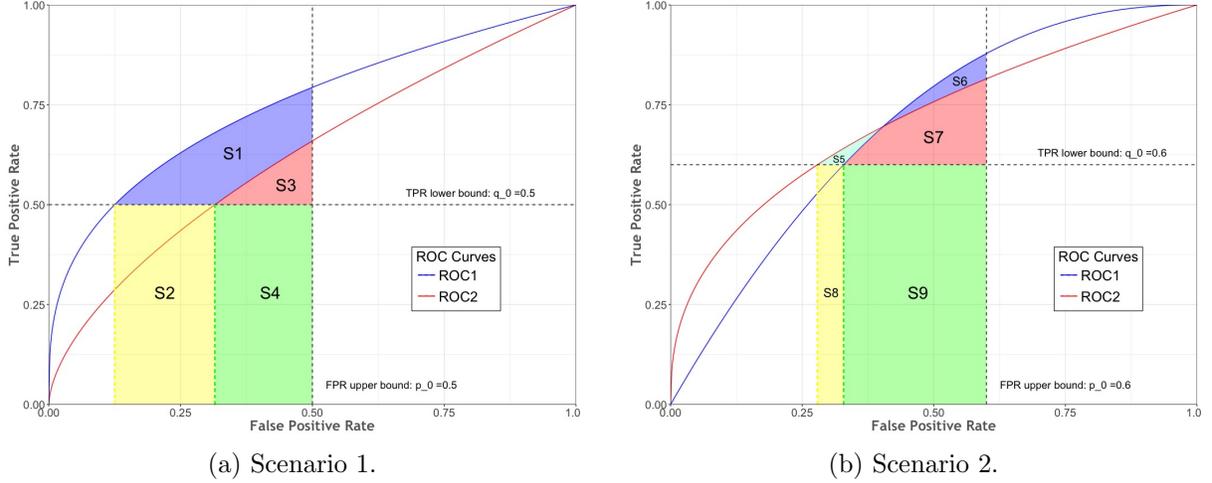

(a) Scenario 1.     (b) Scenario 2.

Figure 2: In Figure 2a, two-way pAUC directly specifies the areas ($S1 + S3$ and $S3$) of interest (restricted by explicit TPR and FPR bounds). FPR pAUC takes redundant regions ($S2 + S4$ and $S4$) into consideration.
In Figure 2b, two-way pAUC is more accurate than FPR pAUC in discriminating two ROC curves. It correctly selects $ROC_1$ that has dominant performance in the restrict region. However, FPR pAUC leads to an opposite (wrong) selection due to that weighty regions below TPR bound are considered.

To sum up, the redundant area in FPR pAUC makes classifier comparison less efficient in some circumstances (in Figure 2a), and may even lead to wrong results (in Figure 2b). In contrast, two-way pAUC provides improvement in both efficiency and accuracy, as illustrated in the above two scenarios.

# 3 Methodology

## 3.1 Nonparametric Estimation

Let $\mathbf{X}$, $\mathbf{Y}$ be the classifier outputs from diseased and non-diseased subjects respectively, and $F(x)$, $G(y)$ are corresponding distribution functions. For any given threshold $c$, a subject is regarded as positive if its classifier output is larger than $c$. TPR and FPR are defined as $S_F(c) := P(\mathbf{X} > c)$ and $S_G(c) := P(\mathbf{Y} > c)$, respectively. Therefore ROC curve is $ROC(u) := S_F\{S_G^{-1}(u)\}$, where $S_G^{-1}(\cdot)$ is the inverse function of the survival function $S_G(\cdot)$. Denote $\{X_i\}_{i=1}^m$ and $\{Y_j\}_{j=1}^n$ as i.i.d random instances from $F(x)$ and $G(y)$ respectively, together with mutual independency. Let $S_{F,m}(u)$ and $S_{G,n}(v)$ be the empirical survival functions, and $S_{F,m}^{-1}(u) := X_{(\lfloor(1-u)m\rfloor)}$ and $S_{G,n}^{-1}(v) := Y_{(\lfloor(1-v)n\rfloor)}$, where $X_{(i)}, Y_{(j)}$ denote the associated order statistics, and $\lfloor C \rfloor$ stands for the largest integer smaller than $C$.

As shown in Figure 1, given bounds on TPR ($\geq p_0$) and FPR ($\leq q_0$), two-way pAUC equals area $A$. It is formulated as

$$U(p_0, q_0) := (\text{ Area A + Area B }) - \text{Area B}$$
$$= \int_{S_G\{S_F^{-1}(q_0)\}}^{p_0} S_F\{S_G^{-1}(u)\}du - [p_0 - S_G\{S_F^{-1}(q_0)\}]q_0. \quad (3.1)$$

A nonparametric estimator $\hat{U}(p_0, q_0)$ directly suggested from (3.1) is

$$\hat{U}(p_0, q_0) := \int_{G_n^{-1}(1-p_0)}^{F_m^{-1}(1-q_0)} F_m\{F_m^{-1}(1-q_0)\}dG_n(t) - \int_{G_n^{-1}(1-p_0)}^{F_m^{-1}(1-q_0)} F_m(t)dG_n(t). \quad (3.2)$$

The consistency of $\hat{U}(p_0, q_0)$ is established in Theorem 4.1. Alternatively, from a probability perspective, $U(p_0, q_0)$ is equivalent to

$$P\{\mathbf{Y} < \mathbf{X}, \mathbf{X} \leq S_F^{-1}(q_0), \mathbf{Y} \geq S_G^{-1}(p_0)\}. \quad (3.3)$$

In other words, two-way pAUC can be viewed as the probability of diseased subject being ranked higher than non-diseased one by classifier with subjects selected from truncated $F(x)$ and $G(y)$. The truncated distributions are determined by selected economical (FPR) and ethical (TPR) quantiles. A trimmed Mann-Whitney U-statistics estimator directly following (3.3) is

$$\frac{1}{mn} \sum_{i=1}^{m} \sum_{j=1}^{n} V_{i,j}(p_0, q_0), \tag{3.4}$$

where $V_{i,j}(p_0, q_0) := I\{Y_j \leq X_i, X_i \leq S_{F,m}^{-1}(q_0), Y_j \geq S_{G,n}^{-1}(p_0)\}$. Proof of Theorem 4.1 shows that (3.2) and (3.4) are exactly equivalent. In other words,

$$\int_{S_G\{S_F^{-1}(q_0)\}}^{p_0} S_F\{S_G^{-1}(u)\}du - [p_0 - S_G\{S_F^{-1}(q_0)\}]q_0 = \frac{1}{mn} \sum_{i=1}^{m} \sum_{j=1}^{n} V_{i,j}(p_0, q_0).$$

Therefore, investigators can use either (3.2) or (3.4) to estimate $U(p_0, q_0)$ without discrepancy in consistency.

**Remark 3.1.** *Two-way pAUC aggregates the discrimination capability of a binary classifier within a given region directly determined by explicit constraints on both TPR and FPR. Previous works, instead, utilize an synthetic approach, namely FPR pAUC here, to put an indirect lower restriction on TPR via artificially setting a corresponding FPR lower bound (6). In particular, $p_1$ is to indirectly lower bound TPR so that it can be maintained at an acceptable level. Let $q_0$ be the lower constraint of interest on TPR. Likewise, $p_2$ is the pre-specific upper bound on FPR. In order to maintain acceptable TPR, investigators denote $p_1 = ROC^{-1}(q_0) = S_G[S_F^{-1}(q_0)]$ as the lower constraint on FPR. FPR pAUC is calculated as follows,*

$$FPR\ pAUC\ (p_1, p_2) = \int_{S_G[S_F^{-1}(q_0)]}^{p_2} ROC(t)dt.$$

**Remark 3.2.** *Similar to FPR pAUC, there exists another indirect synthetic approach, named as TPR pAUC (11). Under TPR and FPR constraints, its philosophy is to indirectly prevent FPR from being too high via setting an artificial upper bound on TPR. Let $q_0$ and $p_0$ be the bounds of interest on FPR ($\leq p_0$) and TPR ($\geq q_0$), respectively. TPR pAUC is formulated as TPR pAUC $(q_0, q_1(p_0)) := \int_{q_0}^{q_1(p_0)} (1 - ROC^{-1}(t))dt$, where $q_1(p_0) = S_F(S_G^{-1}(p_0))$ is the artificial TPR upper bound determined* by the pre-specific FPR upper bound $p_0$. Similar arguments in Section 2 also apply to TPR pAUC. We eliminate the details.

Comparison of two classifiers is of primary concern in clinical research. The difference of discrimination capability measure is a common index therein. In particular, denote the difference of two-way pAUC as $\theta(p_0, q_0) = U_1(p_0, q_0) - U_2(p_0, q_0)$. We estimate the difference via $\hat{\theta}(p_0, q_0) = \hat{U}_1(p_0, q_0) - \hat{U}_2(p_0, q_0)$, where $\hat{U}_k(p_0, q_0)$ is computed from $\{X_{ki}\}_{i=1}^n$ and $\{Y_{kj}\}_{j=1}^n$, $k \in \{1, 2\}$. If $\hat{\theta}(p_0, q_0)$ is significantly positive or negative, the better classifier could be selected. In practice, two classifiers are usually obtained from the same individuals, and thus correlated (12; 13). In other words, $\{X_{1i}\}_{i=1}^n$ and $\{X_{2i}\}_{i=1}^n$ are dependent, so are $\{Y_{1j}\}_{j=1}^n$ and $\{Y_{2j}\}_{j=1}^n$. Note that $\{X_{ki}\}_{i=1}^n$ and $\{Y_{kj}\}_{j=1}^n$ are mutually independent for any $k \in \{1, 2\}$. Due to their (potentially) complicated correlation structure, bootstrap is commonly utilized to approximate the asymptotic distribution of $\hat{\theta}(p_0, q_0)$ (14; 6). Its bootstrap consistency is provided in Theorem 4.2.

## 3.2  Regression Analysis

In this section, a regression analysis framework is introduced for underlying covariates' effects on the classification performance of two-way pAUC. Let $\mathbf{Z}^d$ and $\mathbf{Z}^{\bar{d}}$ be covariates of interest from bounded spaces. $\mathbf{Z}^d$ and $\mathbf{Z}^{\bar{d}}$ represent the covariates of the diseased and the non-diseased respectively. Hence, observations are $(X_i, \mathbf{Z}_i^d)$, $(Y_j, \mathbf{Z}_j^{\bar{d}})$, where $\{(X_i, \mathbf{Z}_i^d)\}_{i=1}^m$ and $\{(Y_j, \mathbf{Z}_j^{\bar{d}})\}_{j=1}^n$ are i.i.d. respectively. Note that both vectors are mutually independent and follow different distributions. To evaluate covariate effects on two-way pAUC, we first define a covariate-specific version of (3.3),

$$U_{\mathbf{Z}_i^d, \mathbf{Z}_j^{\bar{d}}}(p_0, q_0) = P\{Y_j < X_i, X_i \leq S_F^{-1}(q_0), Y_j \geq S_G^{-1}(p_0) | \mathbf{Z}_i^d, \mathbf{Z}_j^{\bar{d}}\}. \tag{3.5}$$

Compared with (3.3), $U_{\mathbf{Z}_i^d, \mathbf{Z}_j^{\bar{d}}}(p_0, q_0)$ conducts the comparison conditional on diseased population with covariate $\mathbf{Z}_i^d$ and the non-diseased with $\mathbf{Z}_j^{\bar{d}}$. Note $\mathbb{E}\{V_{i,j}(p_0, q_0) | \mathbf{Z}_i^d, \mathbf{Z}_j^{\bar{d}}\} = U_{\mathbf{Z}_i^d, \mathbf{Z}_j^{\bar{d}}}(p_0, q_0)$. Then, we propose the regression model for two-way pAUC

$$U_{\mathbf{Z}_i^d, \mathbf{Z}_j^{\bar{d}}}(p_0, q_0) = \eta(\mathbf{Z}_{i,j}^\top \boldsymbol{\beta}_0),$$

where $\eta$ is the inverse of link function and $\mathbf{Z}_{i,j}$ is the abbreviation for $(\mathbf{Z}_i^d, \mathbf{Z}_j^{\bar{d}})^\top$. One popular choice of link function is logit. In this case, effects of covariates are represented as two-way pAUC odds, $U_{\mathbf{Z}_i^d, \mathbf{Z}_j^{\bar{d}}}(p_0, q_0) / \big((1-q_0)p_0 - U_{\mathbf{Z}_i^d, \mathbf{Z}_j^{\bar{d}}}(p_0, q_0)\big)$. Since odds are monotonic increasing with two-way pAUC, positive elements of $\boldsymbol{\beta}_0$ indicate corresponding covariates have improvement on classification accuracy.

To estimate $\boldsymbol{\beta}_0$, we take differential of log-likelihood and get

$$\begin{aligned}
&S_{m,n}(\boldsymbol{\beta}) \\
&= \frac{1}{mn} \sum_{i=1}^m \sum_{j=1}^n \frac{\partial U_{\mathbf{Z}_i^d, \mathbf{Z}_j^{\bar{d}}}(p_0, q_0)}{\partial \boldsymbol{\beta}} \Big(V_{i,j}(p_0, q_0) - U_{\mathbf{Z}_i^d, \mathbf{Z}_j^{\bar{d}}}(p_0, q_0)\Big) \Big(U_{\mathbf{Z}_i^d, \mathbf{Z}_j^{\bar{d}}}(p_0, q_0)\big(1 - U_{\mathbf{Z}_i^d, \mathbf{Z}_j^{\bar{d}}}(p_0, q_0)\big)\Big)^{-1} \\
&= \frac{1}{mn} \sum_{i=1}^m \sum_{j=1}^n S_{i,j}(\boldsymbol{\beta}),
\end{aligned}$$

where $U_{\mathbf{Z}_i^d, \mathbf{Z}_j^{\bar{d}}}(p_0, q_0)\big(1 - U_{\mathbf{Z}_i^d, \mathbf{Z}_j^{\bar{d}}}(p_0, q_0)\big) \geq c$ for $\boldsymbol{\beta} \in N_\delta(\boldsymbol{\beta}_0) = \{\boldsymbol{\beta} : \|\boldsymbol{\beta} - \boldsymbol{\beta}_0\|_2 < \delta\}$ and some $c > 0$. Estimator $\hat{\boldsymbol{\beta}}$ is the solution of $S_{m,n}(\boldsymbol{\beta}) = 0$. Theorem 4.3 ensures the existence and uniqueness of $\hat{\boldsymbol{\beta}}$. Its asymptotic normality is also derived.

## 4 Main Theorems

In this section, we derive the asymptotic properties of $U(p_0, q_0)$, $\theta(p_0, q_0)$ and $\boldsymbol{\beta}_0$. Detailed proofs are demonstrated in the Supplementary Material.

Asymptotic properties of nonparametric estimator $\hat{U}(p_0, q_0)$ require some conditions as follows:

**Assumption 4.1.** *(i) $F^{-1}(1 - q_0)$ is the unique solution of $F(t-) \leq 1 - q_0 \leq F(t)$, $0 \leq q_0 \leq 1$; (ii) $F(t)$ is differentiable; (iii) $F(t)$ is twice differentiable at $F^{-1}(1-q_0)$; (iv) $F'\{F^{-1}(1-q_0)\} > 0$.*

**Assumption 4.2.** *(i) $G^{-1}(1 - p_0)$ is the unique solution of $G(t-) \leq 1 - p_0 \leq G(t)$, $0 \leq p_0 \leq 1$; (ii) $G(t)$ is differentiable; (iii) $G(t)$ is twice differentiable at $G^{-1}(1-p_0)$; (iv) $G'\{G^{-1}(1-p_0)\} > 0$.*

**Theorem 4.1.** *Under Assumption 4.1 and 4.2, we have*

$$\sqrt{m+n}\{\hat{U}(p_0,q_0) - U(p_0,q_0)\} \xrightarrow{d} N\left\{0, \frac{\sigma_3^2}{\lambda} + \frac{\sigma_4^2}{1-\lambda}\right\}, \quad as \quad m, n \to \infty,$$

where $\frac{m}{m+n} \to \lambda$,

$$\sigma_3^2 = F\{G^{-1}(1-p_0)\}[G\{F^{-1}(1-q_0)\} - (1-p_0)]^2 + \int_{G^{-1}(1-p_0)}^{F^{-1}(1-q_0)} [G\{F^{-1}(1-q_0)\} - G(t)]^2 dF(t)$$

$$- \left\{\int_{G^{-1}(1-p_0)}^{F^{-1}(1-q_0)} F(t) dG(t)\right\}^2,$$

and

$$\sigma_4^2 = [1 - q_0 - F\{G^{-1}(1-p_0)\}]^2(1-p_0) + \int_{G^{-1}(1-p_0)}^{F^{-1}(1-q_0)} \{1 - q_0 - F(t)\}^2 dG(t)$$

$$- \left\{\int_{G^{-1}(1-p_0)}^{F^{-1}(1-q_0)} G(t) dF(t)\right\}^2.$$

From Theorem 4.1, the asymptotic $100(1-\alpha)\%$ confidence interval for $U(p_0, q_0)$ is

$$\left(\hat{U}(p_0, q_0) - \frac{Z_{1-\alpha/2}}{\sqrt{m+n}}\sqrt{\frac{\sigma_3^2}{\lambda} + \frac{\sigma_4^2}{1-\lambda}},\ \hat{U}(p_0, q_0) + \frac{Z_{1-\alpha/2}}{\sqrt{m+n}}\sqrt{\frac{\sigma_3^2}{\lambda} + \frac{\sigma_4^2}{1-\lambda}}\right), \quad (4.1)$$

where $Z_{1-\alpha/2}$ stands for $1 - \alpha/2$ quantile of standard normal distribution.

To conduct inference for the difference of two correlated pAUCs, we use the following bootstrap method. Let us bootstrap resample (uniformly with replacement) $B$ times from the given sample and calculate $B$ bootstrap estimates $\{\hat{\theta}_i(p_0, q_0)\}_{i=1}^B$ (of $\hat{\theta}(p_0, q_0)$). Bootstrap variance estimator is $v_{boot}^2(p_0, q_0) = 1/B \sum_{i=1}^B (\hat{\theta}_i(p_0, q_0) - 1/B \sum_{r=1}^B \hat{\theta}_r(p_0, q_0))^2$. Asymptotic normality of $(\hat{\theta}(p_0, q_0) - \theta(p_0, q_0))/v_{boot}$ is established in Theorem 4.2.

**Theorem 4.2.** *Under the same conditions of Theorem 4.1, we have*

$$\sqrt{m+n}\left(\frac{\hat{\theta}(p_0, q_0) - \theta(p_0, q_0)}{v_{boot}(p_0, q_0)}\right) \xrightarrow{d} N(0, 1), \quad as \quad m, n, B \to \infty.$$

From Theorem 4.2, a $100(1-\alpha)\%$ confidence interval of $\theta$ is

$$\left(\hat{\theta}(p_0, q_0) - \frac{Z_{1-\alpha/2}}{\sqrt{m+n}} v_{boot}(p_0, q_0),\ \hat{\theta}(p_0, q_0) + \frac{Z_{1-\alpha/2}}{\sqrt{m+n}} v_{boot}(p_0, q_0)\right). \tag{4.2}$$

As the next step of comparison, the relationship between classification accuracy and underlying factors is of further interest. To theoretically justify the regression framework we proposed in Section 3.2, two different conditions are proposed as follows.

**Assumption 4.3.** $\eta(\cdot)$ is three-times differentiable with bounded derivatives and monotonic increasing.

**Assumption 4.4.** Matrix $\mathbb{E}\{\partial S_{i,j}(\boldsymbol{\beta}_0)/\partial \boldsymbol{\beta}\}$ is negative definite.

To ensure the validity of the regression framework, the estimator $\hat{\boldsymbol{\beta}}$ proposed in Section 3.2 must uniquely exist and be consistent to the true parameter $\boldsymbol{\beta}_0$. Theorem 4.3 ensures the uniqueness and consistency of $\hat{\boldsymbol{\beta}}$.

**Theorem 4.3.** *Under Assumption 4.3 and 4.4, as $m, n \to \infty$, there exists a unique solution $\hat{\boldsymbol{\beta}}$ to $S_{m,n}(\boldsymbol{\beta}) = 0$ with probability converging to one, and*

$$\hat{\boldsymbol{\beta}} \xrightarrow{p} \boldsymbol{\beta}_0.$$

Moreover, the asymptotic distribution of $\hat{\beta}$ is of certain interest. To simplify notation, we define

$$\boldsymbol{\omega}_{i,j} = \frac{\partial U_{\mathbf{Z}_i^d, \mathbf{Z}_j^{\bar{d}}}(p_0, q_0)}{\partial \boldsymbol{\beta}} \left(U_{\mathbf{Z}_i^d, \mathbf{Z}_j^{\bar{d}}}(p_0, q_0)\bigl(1 - U_{\mathbf{Z}_i^d, \mathbf{Z}_j^{\bar{d}}}(p_0, q_0)\bigr)\right)^{-1},$$

$$\mathbb{E}\{V_{i,j}(p_0, q_0)|\mathbf{Z}_j^{\bar{d}}\} = G_{\mathbf{Z}_j^{\bar{d}}}(X_i),\ \mathbb{E}\{V_{i,j}(p_0, q_0)|\mathbf{Z}_i^d\} = S_{F,\mathbf{Z}_i^d}(Y_j).$$

The following theorem establishes the asymptotic normality of $\hat{\boldsymbol{\beta}}$.

**Theorem 4.4.** *Under the same conditions of Theorem 4.3, as $m, n \to \infty$,*

$$\sqrt{\frac{mn}{m+n}}(\hat{\boldsymbol{\beta}} - \boldsymbol{\beta}_0) \xrightarrow{d} N\left(\mathbf{0}, \boldsymbol{\Delta}\bigl[(1-\lambda)\boldsymbol{\Sigma}_X + \lambda\boldsymbol{\Sigma}_Y\bigr]\boldsymbol{\Delta}\right),$$

where $\mathbf{\Delta} = \left(\mathbb{E}\{\partial S_{i,j}(\boldsymbol{\beta}_0)/\partial \boldsymbol{\beta}\}\right)^{-1}$, $\frac{m}{m+n} \to \lambda$,

$$\boldsymbol{\Sigma}_X = \lim_{m,n\to\infty} \left[\frac{1}{m}\sum_{i=1}^{m}\frac{1}{n^2}\sum_{j=1}^{n}\sum_{l=1}^{n}\boldsymbol{\omega}_{i,j}\boldsymbol{\omega}_{i,l}^T Cov\bigg(G_{\mathbf{Z}_j^{\bar{d}}}(X_i), G_{\mathbf{Z}_l^d}(X_i)\bigg)\right], \text{ and}$$

$$\boldsymbol{\Sigma}_Y = \lim_{m,n\to\infty} \left[\frac{1}{n}\sum_{j=1}^{n}\frac{1}{m^2}\sum_{i=1}^{m}\sum_{k=1}^{m}\boldsymbol{\omega}_{i,j}\boldsymbol{\omega}_{k,j}^T Cov\bigg(S_{F,\mathbf{Z}_i^d}(Y_j), S_{F,\mathbf{Z}_k^d}(Y_j)\bigg)\right].$$

Therefore, inferences for $\beta$ can be conduct by the asymptotic distribution of $\sqrt{\frac{mn}{m+n}}(\hat{\boldsymbol{\beta}} - \boldsymbol{\beta}_0)$, which is a normal distribution with mean 0 and variance $\Delta[(1-\lambda)\Sigma_X + \lambda\Sigma_Y]\Delta$. To obtain asymptotic variance in Theorem 4.4, $G_{\mathbf{Z}_j^{\bar{d}}}(X_i)$ and $S_{F,\mathbf{Z}_i^d}(Y_j)$ can be estimated empirically. For discrete covariates $\mathbf{Z}_i^d$ and $\mathbf{Z}_j^{\bar{d}}$ with sufficient large sample sizes at each level, simple average can be applied to estimate conditional expectations. For continuous covariates, the bootstrap method is recommended in (3).

## 5 Numerical Study

In this section, numerical study for the above methods are conducted. Additional simulation results are shown in the Supplementary Material. We encodes the methods as a publicly available R package **tpAUC**, and implement the package for the numerical study.

*Case 1: Asymptotic Normality of Estimators.* We study coverage probability of confidence interval (4.1) to support the asymptotic normality of $\hat{U}(p_0, q_0)$ in Theorem 4.1. FPR ($\leq p_0$) and TPR ($\geq q_0$) constraints are $(p_0, q_0) = (0.6, 0.4)$. Sample sizes $(m, n)$ are chosen as: $(30, 30)$, $(50, 50)$, $(80, 80)$, $(100, 100)$, $(150, 100)$, $(150, 150)$, $(200, 150)$, and $(200, 200)$. In dataset A, **X** and **Y** are generated from $N(1, 1)$ and $N(0, 1)$ respectively; in dataset B, **Y** follows $N(0, 1)$, and **X** follows Exp(1); in dataset C, **X** follows Exp(1) and **Y** follows Exp(0.5). We repeat 1000 times in each setting. As shown in Table 1, the converge probabilities get closer to 95% as sample size growing. These around 95% coverage probabilities support Theorem 4.1.

Table 1: Coverage probability (CP) of 95% confidence interval (4.1) for $\hat{U}(p_0, q_0)$ with $p_0 = 0.6, q_0 = 0.4$ in data set A, B, and C, respectively.

| $m$ | $n$ | CP A | CP B | CP C |
|---|---|---|---|---|
| 30 | 30 | 0.907 | 0.924 | 0.892 |
| 50 | 50 | 0.918 | 0.927 | 0.905 |
| 80 | 80 | 0.923 | 0.937 | 0.925 |
| 100 | 100 | 0.938 | 0.934 | 0.934 |
| 150 | 100 | 0.953 | 0.952 | 0.954 |
| 150 | 150 | 0.934 | 0.947 | 0.937 |
| 200 | 150 | 0.948 | 0.947 | 0.949 |
| 200 | 200 | 0.941 | 0.947 | 0.946 |

Note: The region of interest is determined by FPR $\leq p_0$ and TPR $\geq q_0$. CP being closer to 95% suggests that the asymptotic normality of $\hat{U}(p_0, q_0)$ in Theorem 4.1 holds.

*Case 2: Bootstrap Property of the Difference.* To justify the argument that two-way pAUC improves the efficiency of ROC curve comparison in Scenario 1 of Section 2, we compare two-way pAUC and FPR pAUC via statistical power. Recall that $\theta(p_0, q_0) = U_1(p_0, q_0) - U_2(p_0, q_0)$. Our null hypothesis is $H_0 : \theta(p_0, q_0) = 0$ and alternative is $H_1 : \theta(p_0, q_0) \neq 0$. Critical values are calculated by Theorem 4.2. If power is higher, the corresponding measure is more powerful (aka, more efficient) in classifier comparison. A popular R package **pROC** is utilized to compute AUCs and FPR pAUCs. Note that **pROC** can only compare FPR pAUCs on the same FPR range. However, given the same TPR constraint, different ROC curves intersect the TPR constraint at distinct FPRs (see Figure 2). Thus, FPR pAUCs on different ROC curves do not share FPR range. To by-pass such incovenience, we first use **pROC** to compare (two curves') FRP pAUCs on the FPR range of $ROC_1$ (corresponding power is P-pROC$_1$). Then, we compare (two curves') FRP pAUCs on the FPR range of $ROC_2$ (corresponding power is P-pROC$_2$). Motivated by Theorem 4.2, we utilize bootstrap to compare FPR pAUCs on different FPR ranges (corresponding power is P-FPR). Bootstrap repetition is 1000. FPR ($\leq p_0$) and TPR ($\geq q_0$) constraints $(p_0, q_0)$ are $(0.5, 0.5)$. Sample sizes $m$ and $n$ are chosen as: $(30, 30)$, $(50, 30)$, $(50, 50)$, $(80, 50)$, $(80, 80)$, $(80, 100)$, and $(100, 80)$. For $ROC_1$ in Figure 2a, **X** follows $N(1, 1)$ and **Y** follows $N(-0.4, 1)$; for $ROC_2$, **X** follows $N(0.3, 1)$ and **Y** follows $N(-0.5, 1)$. Let significant level be 0.05. We repeat 1000 times in each setting. As shown in Table 2, two-way pAUC apparently has the highest power in all settings.

This phenomenon supports the argument in Section 2 that FPR pAUC is less efficient (aka, less powerful) than two-way pAUC due to the redundant area below TPR constraint.

Table 2: Power for ROC curves comparison by AUC, two-way pAUC, and FPR pAUC.

| $m$ | $n$ | AUC | | | Two-way pAUC | | | FPR pAUC | | | | |
|---|---|---|---|---|---|---|---|---|---|---|---|---|
| | | $ROC_1$ | $ROC_2$ | P-AUC | $ROC_1$ | $ROC_2$ | P-TW | $ROC_1$ | $ROC_2$ | P-FPR | P-pROC$_1$ | P-pROC$_2$ |
| 30 | 30 | 0.8393 | 0.714 | 0.329 | 0.1142 | 0.0442 | 0.92 | 0.3236 | 0.1885 | 0.916 | 0.317 | 0.276 |
| 50 | 30 | 0.8383 | 0.716 | 0.384 | 0.1145 | 0.0459 | 0.934 | 0.3221 | 0.1902 | 0.928 | 0.366 | 0.307 |
| 50 | 50 | 0.8396 | 0.7134 | 0.496 | 0.113 | 0.0458 | 0.976 | 0.324 | 0.1891 | 0.967 | 0.458 | 0.409 |
| 80 | 50 | 0.8391 | 0.7142 | 0.585 | 0.1158 | 0.046 | 0.993 | 0.3235 | 0.1896 | 0.982 | 0.533 | 0.498 |
| 80 | 80 | 0.8386 | 0.714 | 0.694 | 0.1147 | 0.0465 | 0.994 | 0.3238 | 0.1895 | 0.99 | 0.644 | 0.595 |
| 80 | 100 | 0.8372 | 0.7122 | 0.746 | 0.1158 | 0.0457 | 0.996 | 0.3229 | 0.1887 | 0.994 | 0.701 | 0.653 |
| 100 | 80 | 0.8389 | 0.7133 | 0.739 | 0.1156 | 0.0462 | 0.998 | 0.3238 | 0.1892 | 0.995 | 0.685 | 0.648 |

Note: The region of interest is determined by FPR $\leq 0.5$ and TPR $\geq 0.5$. P-AUC and P-TW denote power by AUC and two-way pAUC, respectively. P-FPR indicates the power by FPR pAUCs on the ranges of $ROC_1$ and $ROC_2$. P-pROC$_1$ is the power by R package **pROC** on the FPR range of $ROC_1$; likewise, P-pROC$_2$ is the power by R package **pROC** on the FPR range of $ROC_2$. As shown in Table 2, two-way pAUC apparently has the highest power in all settings.

*Case 3: Type* **I** *Error when explicit restricted regions are small.* We study type **I** errors of classifier comparison by two-way pAUC and FPR pAUC respectively. In particular, we are interested in the effect of two-way pAUC's size on its type **I** error. Null hypothesis is $H_0 : \theta_0(p_0, q_0) = 0$. For two ROC curves, the diseased observations are both generated from $N(1,1)$, and the non-diseased are both from $N(0,1)$. Let significant level be 0.05. Simulation and bootstrap both repeat 1000 times. As shown in Table 3, type **I** error gets closer to 0.05 as sample size grows. In addition, the larger size of two-way pAUC is, the closer its type **I** error is to 0.05. We believe that this phenomenon could be interpreted by the fact that larger number of active observations, which fall into the restricted region, essentially benefit statistical properties, e.g., type I error.

# 6 Application in Breast Cancer Data

In this section, we apply the above methods into Wisconsin Breast Cancer Data (Diagnostic). This dataset records diagnosis results of breast cancer, in which 30 biomarkers are measured from 469 subjects (189 malignant and 280 benign). To compare classification performance, we focus on two

Table 3: Type **I** errors of Two-way (TW) pAUC and FPR pAUC.

| $p_0$ | $q_0$ | $m$ | $n$ | TW | TW Type **I** | FPR | FPR Type **I** |
|---|---|---|---|---|---|---|---|
| 0.3 | 0.5 | 50 | 50 | 0.014 | 0.034 | 0.085 | 0.041 |
| 0.3 | 0.5 | 100 | 100 | 0.014 | 0.048 | 0.085 | 0.041 |
| 0.3 | 0.5 | 200 | 200 | 0.014 | 0.039 | 0.085 | 0.042 |
| 0.4 | 0.5 | 50 | 50 | 0.037 | 0.043 | 0.156 | 0.029 |
| 0.4 | 0.5 | 100 | 100 | 0.037 | 0.034 | 0.157 | 0.028 |
| 0.4 | 0.5 | 200 | 200 | 0.037 | 0.043 | 0.158 | 0.041 |
| 0.5 | 0.5 | 50 | 50 | 0.067 | 0.039 | 0.238 | 0.028 |
| 0.5 | 0.5 | 100 | 100 | 0.068 | 0.033 | 0.239 | 0.029 |
| 0.5 | 0.5 | 200 | 200 | 0.068 | 0.050 | 0.238 | 0.038 |
| 0.4 | 0.6 | 50 | 50 | 0.017 | 0.030 | 0.120 | 0.039 |
| 0.4 | 0.6 | 100 | 100 | 0.016 | 0.033 | 0.119 | 0.044 |
| 0.4 | 0.6 | 200 | 200 | 0.016 | 0.048 | 0.119 | 0.054 |
| 0.5 | 0.6 | 50 | 50 | 0.036 | 0.043 | 0.199 | 0.033 |
| 0.5 | 0.6 | 100 | 100 | 0.037 | 0.035 | 0.200 | 0.039 |
| 0.5 | 0.6 | 200 | 200 | 0.037 | 0.044 | 0.200 | 0.040 |
| 0.4 | 0.7 | 50 | 50 | 0.003 | 0.012 | 0.061 | 0.028 |
| 0.4 | 0.7 | 100 | 100 | 0.003 | 0.011 | 0.061 | 0.041 |
| 0.4 | 0.7 | 200 | 200 | 0.003 | 0.013 | 0.061 | 0.044 |
| 0.5 | 0.7 | 50 | 50 | 0.014 | 0.024 | 0.142 | 0.036 |
| 0.5 | 0.7 | 100 | 100 | 0.013 | 0.030 | 0.142 | 0.042 |
| 0.5 | 0.7 | 200 | 200 | 0.014 | 0.038 | 0.142 | 0.045 |

Note: The region of interest is determined by FPR $\leq p_0$ and TPR $\geq q_0$. TW and TW Type **I** denote average value of estimations and type **I** error of two-way pAUC, respectively. Likewise, FPR and FPR Type **I** are average value of estimations and type **I** error of FPR pAUC, respectively. As shown in the table, two-way pAUC's type **I** error gets closer to 0.05 as either sample size grows or its size increases.

markers: Concavity SE and Worst Smoothness.

Breast cancer is a lethal disease. If malignant subjects cannot be identified, ethical consequences will be serious. Therefore, we set FPR ($\leq p_0$) and TPR ($\geq q_0$) constraints as $(p_0, q_0) = (0.35, 0.5)$. ROC curves for the two markers are in Figure 3. FPR lower bounds in FPR pAUCs for Worst Smoothness and Concavity SE are 0.152 and 0.19, respectively. As shown in Figure 3, it clear that the Concavity SE has better discrimination capability than Worst Smoothness, since the former has larger area under its ROC curve in the region of interest. In order to distinguish comparison accuracy of the performance measures, we utilize these measures to compare the two markers. Package **pROC** is applied to estimate FPR pAUC. Table 4 demonstrates that only two-way pAUC selects Concavity SE as the better classifier. However, due to a redundant region S is taken into consideration, FPR pAUC leads to the opposite selection. This phenomenon supports the argument in Scenario 2 of Section 2 that FPR pAUC is less accurate in classifier comparison than two-way pAUC.

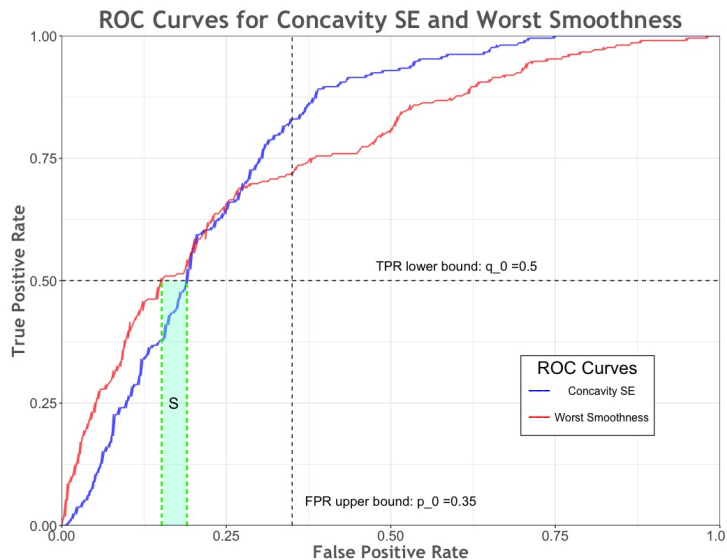

Figure 3: ROC curves for two markers: Concavity SE and Worst Smoothness. Combining Figure 3 and Table 4, given the region of interest (TPR $\geq 0.5$ and FPR $\leq 0.35$), two-way pAUC selects the marker that has better classification performance, i.e., Concavity SE. Whereas FPR pAUC leads to the opposite selection, due to the incorporation of the redundant (shaded) region S.

Then, we turn to regression analysis of two-way pAUC and FPR pAUC. The following empirical

Table 4: Estimates of three performance measures given the region of interest (TPR $\geq 0.5$ and FPR $\leq 0.35$).

| Biomarker | AUC | FPR pAUC | Two-way pAUC |
|---|---|---|---|
| Concavity SE | 0.7808 | 0.1101 | 0.0311 |
| Worst Smoothness | 0.7541 | 0.1253 | 0.0278 |

Note: FPR lower bounds in FPR pAUCs for Worst Smoothness and Concavity SE are 0.152 and 0.19, respectively. Larger value of certain measure suggests that the corresponding marker has better classification performance in terms of this measure. Combining Figure 3 and Table 4, only two-way pAUC selects Worst Smoothness, the marker enjoys better performance in the restrict region.

study shows that investigators may be misled by FPR pAUC regression (6) as well. To be specific, the model of two-way pAUC regression is

$$\log\left\{\frac{U_{\mathbf{Z}_i^d, \mathbf{Z}_j^{\bar{d}}}(p_0, q_0)}{(1-q_0)p_0 - U_{\mathbf{Z}_i^d, \mathbf{Z}_j^{\bar{d}}}(p_0, q_0)}\right\} = \beta_0 + \beta_1 Z_{j1}^{\bar{d}} + \beta_2 Z_{j2}^{\bar{d}} + \beta_3 Z_{i1}^d + \beta_4 Z_{i2}^d,$$

where $Z_{j1}^{\bar{d}}, Z_{j2}^{\bar{d}}, Z_{i1}^d$, and $Z_{i2}^d$ are non-diseased compactness SE, non-diseased concavity SE, diseased compactness SE, and diseased concavity SE respectively. FPR and TPR constraints are $(p_0, q_0)$. Moreover, we compare two-way pAUC regression with FPR pAUC regression under the same setting. For FPR pAUC regression, the range of FPR is $(S_G(S_F^{-1}(q_0)), p_0)$. As shown in Table 5, $\beta_3$ is negative in two-way pAUC regression while positive in FPR pAUC regression. It concludes that $Z_{i1}^d$ is in negative relation with classification accuracy in the region of interest. However, due to the redundant area in FPR pAUC, the true relation is distorted.

Table 5: Two-way partial AUC (TW) v.s. FPR partial AUC (FPR) in regression parameters estimation using Wisconsin Breast Cancer Data.

| $p_0$ | $q_0$ | $\beta_0$ | | $\beta_1$ | | $\beta_2$ | | $\beta_3$ | | $\beta_4$ | |
|---|---|---|---|---|---|---|---|---|---|---|---|
| | | TW | FPR | TW | FPR | TW | FPR | TW | FPR | TW | FPR |
| 0.5 | 0.5 | -1.04 | -1.94 | -0.20 | 6.25 | -6.01 | -6.93 | -32.89 | 13.24 | -11.76 | 6.63 |
| 0.5 | 0.6 | -1.13 | -1.86 | -2.23 | -1.30 | -7.06 | -5.05 | -52.96 | 11.15 | -9.36 | 5.68 |
| 0.6 | 0.5 | -0.13 | -1.24 | -13.52 | -7.01 | -4.19 | -5.53 | -40.77 | 13.53 | -12.41 | 6.90 |
| 0.6 | 0.6 | -0.14 | -1.14 | -16.97 | -15.09 | -4.56 | -3.53 | -63.25 | 11.29 | -9.41 | 5.97 |

Note: The region of interest is determined by FPR $\leq p_0$ and TPR $\geq q_0$. As shown in the table, both $\beta_3$ and $\beta_4$ are in negative relation with the area under the ROC curve in the restricted region (by two-way pAUC regression). However, FPR pAUC regression leads to opposite relation due to the distortion of redundant area (with TPR $\leq q_0$).

# 7 Discussion

In this paper, we propose a novel performance measure, named as two-way pAUC, which directly quantifies the discrimination capability of the restricted region under ROC curve simultaneously determined by explicit TPR and FPR constraints. Previous sections demonstrate that, compared with existing FPR pAUC, two-way pAUC has obvious advantages in implementation and analysis. In particular, FPR pAUC is not able to directly control TPR since the area below TPR lower constraint is always considered. In addition, such redundant area decreases its efficiency and accuracy (link to statistical power) for ROC curve comparison (refer to Section 2, 5 & 6). Therefore, considering explicit TPR and FPR restrictions, two-way pAUC is a more practical tool for ROC analysis.

In practice, TPR and FPR constraints should be carefully selected. For clinical needs, high TPR lower bound and low FPR upper bound are preferred. In this case, ethical and economical consequences can be controlled. In contrast, from a statistical perspective, low TPR lower bound and high FPR upper bound would be better. Under this circumstance, more observations would fall into the (larger) restricted region. Thus, statistical properties, e.g., power and type I error, are ensured. We recommend investigators to find a compromise between clinical needs and statistical properties in the design phase of the study. In particular, a pilot study or priori knowledge may be helpful to select appropriate TPR and FPR constraints, and determine corresponding necessary sample size.

*Supplementary Material to*

# Two-Way Partial AUC and Its Properties

Hanfang Yang[*]   Kun Lu[†]   Xiang Lyu[‡]   Feifang Hu[§]

The supplementary material is organized as follows:

- In Section S.1, we provide key lemmas to the proof of main results.

- In Section S.2, we prove Theorem 4.1.

- In Section S.3, we prove Theorem 4.2.

- In Section S.4, we prove Theorem 4.3.

- In Section S.5, we prove Theorem 4.4.

- In Section S.6, we provide additional simulation results.

## S.1   Key Lemmas

**Lemma S.1.1.** *Let $m$ and $n$ be sequences of integers such that $\frac{m}{m+n} \to \lambda$, $0 < \lambda < 1$, as $m, n \to \infty$; $F(t)$, $G(t)$ be continuous; $F^{-1}(1-q_0)$ be the unique solution of $F(-t) < 1 - q_0 < F(t)$, $0 < q_0 < 1$. Then,*

$$\sqrt{m+n} \int_{F^{-1}(1-q_0)}^{F_m^{-1}(1-q_0)} \{F_m(t) - F(t)\} dG(t) = o_p(1), \quad m, n \to \infty. \tag{S.1}$$

[*]Corresponding Author. Associate Professor, School of Statistics, Renmin University, China. E-mail: hyang@ruc.edu.cn.
[†]PhD Student, Department of Operations Research and Financial Engineering, Princeton University, Princeton, NJ, USA.
[‡]PhD Student, Department of Statistics, Purdue University, West Lafayette, IN, USA.
[§]Professor, Department of Statistics, George Washington University, Washington, D.C., USA.

*Proof.* We can easily see that,

$$\left|\sqrt{m+n}\int_{F^{-1}(1-q_0)}^{F_m^{-1}(1-q_0)}\{F_m(t)-F(t)\}dG(t)\right| \leq \left|\int_{F^{-1}(1-q_0)}^{F_m^{-1}(1-q_0)}\sqrt{m+n}\sup_t|F_m(t)-F(t)|dG(t)\right|$$

$$= \sup_t \sqrt{m+n}|F_m(t)-F(t)|\left|\int_{F^{-1}(1-q_0)}^{F_m^{-1}(1-q_0)}dG(t)\right| \quad \text{(S.2)}$$

Firstly,

$$\sup_t \sqrt{m+n}|F_m(t)-F(t)| = \sqrt{\frac{m+n}{m}}\sup_t \sqrt{m}|F_m(t)-F(t)|.$$

Because

$$\sup_t \sqrt{m}|F_m(t)-F(t)| = \mathbf{O}_p(1),$$

and

$$\sqrt{\frac{m+n}{m}} \to \sqrt{\frac{1}{\lambda}}, \quad m,n \to \infty.$$

Therefore

$$\sup_t \sqrt{m+n}|F_m(t)-F(t)| = \mathbf{O}_p(1), \quad m,n \to \infty. \quad \text{(S.3)}$$

Then we consider term $\left|\int_{F^{-1}(1-q_0)}^{F_m^{-1}(1-q_0)}dG(t)\right|$.

$$\int_{F^{-1}(1-q_0)}^{F_m^{-1}(1-q_0)}dG(t) = G\{F_m^{-1}(1-q_0)\} - G\{F^{-1}(1-q_0)\}$$

$$= G\{F^{-1}(1-q_0)\} - G_n\{F^{-1}(1-q_0)\} + \mathbf{O}(n^{-1})$$

$$= o(1), \quad m \to \infty. \quad \text{(S.4)}$$

Then apply (S.2) and (S.3) to (S.4), we have

$$\sqrt{m+n}\int_{F^{-1}(1-q_0)}^{F_m^{-1}(1-q_0)}\{F_m(t)-F(t)\}dG(t) = o_p(1), \quad m,n \to \infty.$$

□

**Lemma S.1.2.** *Let $F(t)$ and $G(t)$ be distribution functions with sample distribution functions $F_m(t)$*

and $G_n(t)$, respectively; $t \in \mathbf{R}$. Let $F(t)$ be continuous. Let $F^{-1}(1 - q_0)$ be the unique solution of $F(t) \leq 1 - q_0 \leq F(t)$, $0 < 1 - q_0 < 1$. Define $f_m(t) = F_m\{F_m^{-1}(1 - q_0)\}I\{t \leq F_m^{-1}(1 - q_0)\}$, and $f_0 = F\{F^{-1}(1 - q_0)\}I\{t \leq F^{-1}(1 - q_0)\}$. Then,

$$\sqrt{m+n}(P_n f_m - P f_m - P_n f_0 + P f_0) = o_p(1), \quad m, n \to \infty.$$

*Equivalently,*

$$\sqrt{m+n}(P_n f_m - P f_0) = \sqrt{m+n}(P_n f_0 - P f_0) + \sqrt{m+n}(P f_n - P f_0) + o_p(1), \quad m, n \to \infty. \quad (S.5)$$

*Proof.* Let us consider the term **I** first. Since $G_n(x) = P_n((-\infty, x])$. In this case, the empirical process is indexed by a class $\mathcal{C} = \{(-\infty, x] : x \in \mathbf{R}\}$, with only one element in this class. It has been shown that $\mathcal{C}$ is a Donsker class, because $\sqrt{m+n}\{G_n(x) - G(x)\}$ converges weakly in $\mathcal{L}^\infty(\mathbf{R})$ to a Brownian bridge $B\{G(x)\}$. Thus, it is not difficult to conclude that $\mathcal{D} = \{F\{F^{-1}(1 - q_0)\}I\{t \leq x\} : x \in \mathbf{R}\}$ is also a Donsker class. Thus for $f_m(t) = F_m\{F_m^{-1}(1 - q_0)\}I\{t \leq F_m^{-1}(1 - q_0)\}$, and $f_0 = F\{F^{-1}(1 - q_0)\}I\{t \leq F^{-1}(1 - q_0)\}$, then they are in class $\mathcal{D}$.

Otherwise,

$$F_m^{-1}(1 - q_0) \to F^{-1}(1 - q_0) \quad wp1, \quad m \to \infty.$$

Then

$$\int [I\{t \leq F_m^{-1}(1 - q_0)\} - I\{t \leq F^{-1}(1 - q_0)\}]dP \xrightarrow{P} 0, \quad m \to \infty.$$

Also note that $\sup_t |H_m(t) - H(t)| \xrightarrow{P} 0$, $m \to \infty$. Thus,

$$\int [f_m(t) - f(t)]^2 dP \xrightarrow{P} 0, \quad m \to \infty.$$

where H(t) is any distribution function.

Therefore by lemma 2.3 in (15), we can easily get

$$\sqrt{m+n}(P_n f_m - P f_m - P_n f_0 + P f_0) = o_p(1), \quad m, n \to \infty.$$

Equally,

$$\sqrt{m+n}(P_n f_m - P f_0) = \sqrt{m+n}(P_n f_0 - P f_0) + \sqrt{m+n}(P f_n - P f_0) + o_p(1), \quad m, n \to \infty.$$

$\square$

**Lemma S.1.3.** *Let $m$ and $n$ be sequences of integers such that $\frac{m}{m+n} \to \lambda$, $0 < \lambda < 1$, as $m, n \to \infty$; $F(t)$, $G(t)$ be continuous; $F^{-1}(1-q_0)$ be the unique solution of $F(-t) < 1 - q_0 < F(t)$, $0 < q_0 < 1$. Then,*

$$\sqrt{m+n} \int_{F^{-1}(1-q_0)}^{F_m^{-1}(1-q_0)} [F_m\{F_m^{-1}(1-q_0)\} - (1-q_0)] dG(t) = o_p(1), m, n \to \infty. \quad (S.6)$$

*Proof.* The proof methods are just exactly the same as Lemma S.1.1. $\square$

**Lemma S.1.4.** *Let $G(t)$ be differentiable; and $G(t)$ be twice differentiable at $F^{-1}(1-q_0)$ and $G'\{F^{-1}(1-q_0)\} > 0$. Then,*

$$\sqrt{m+n} \left\{ \int_{-\infty}^{F_m^{-1}(1-q_0)} F(t) dG(t) - \int_{-\infty}^{F^{-1}(1-q_0)} F(t) dG(t) \right\}$$
$$= \sqrt{m+n} \frac{[1 - q_0 - F_m\{F^{-1}(1-q_0)\}]}{F'\{F^{-1}(1-q_0)\}} (1-q_0) G'\{F^{-1}(1-q_0)\} + o_p(1), \quad m \to \infty. \quad (S.7)$$

*Proof.* By Taylor Expansion, we have

$$\sqrt{m+n} \left\{ \int_{-\infty}^{F_m^{-1}(1-q_0)} F(t) dG(t) - \int_{-\infty}^{F^{-1}(1-q_0)} F(t) dG(t) \right\}$$
$$= \sqrt{m+n} \{F_m^{-1}(1-q_0) - F^{-1}(1-q_0)\} F\{F^{-1}(1-q_0)\} G'\{F^{-1}(1-q_0)\} + o_p(1), \quad m \to \infty.$$

Then by Theorem 2.13 in (15), we have

$$\sqrt{m}[F_m^{-1}(1-q_0) - F^{-1}(1-q_0)] = \sqrt{m} \frac{1 - q_0 - F_m\{F^{-1}(1-q_0)\}}{F'\{F^{-1}(1-q_0)\}} + o_p(1), \quad n \to \infty.$$

Thus

$$\sqrt{m+n}\left\{\int_{-\infty}^{F_m^{-1}(1-q_0)} F(t)dG(t) - \int_{-\infty}^{F^{-1}(1-q_0)} F(t)dG(t)\right\}$$
$$=\sqrt{m+n}\frac{1-q_0-F_m\{F^{-1}(1-q_0)\}}{F'\{F^{-1}(1-q_0)\}}(1-q_0)G'\{F^{-1}(1-q_0)\}+o_p(1), \quad m\to\infty.$$

□

**Lemma S.1.5.** *Let $G(t)$ be differentiable; and $G(t)$ be twice differentiable at $F^{-1}(1-q_0)$ and $G'\{F^{-1}(1-q_0)\}>0$. Then,*

$$\sqrt{m+n}\left\{\int_{-\infty}^{F_m^{-1}(1-q_0)}(1-q_0)dG(t)-\int_{-\infty}^{F^{-1}(1-q_0)}(1-q_0)dG(t)\right\}$$
$$=(1-q_0)\sqrt{m+n}\frac{[1-q_0-F_m\{F^{-1}(1-q_0)\}]}{F'\{F^{-1}(1-q_0)\}}G'\{F^{-1}(1-q_0)\}+o_p(1). \tag{S.8}$$

*Proof.* The proof methods are just exactly the same as Lemma S.1.4. □

## S.2 Proof of Theorem 4.1

Proof of this part follows similar steps in (15). The main idea is continuing splitting term $\sqrt{m+n}(\hat{U}-U)$ until it is divided into two parts that only depend on $m$ or $n$ respectively. Here we only provide some major procedures, detailed deduction process can be referred to in the complementary material.

At first, we need to show:

$$\frac{1}{mn}\sum_{i=1}^{m}\sum_{j=1}^{n}I(Y_j \leq X_i)I\{X_i \leq S_{F,m}^{-1}(q_0)\}I\{Y_j \geq S_{G,n}^{-1}(p_0)\}$$

$$=\frac{1}{n}\sum_{j=1}^{n}\frac{1}{m}\sum_{i=1}^{m}I(Y_j \leq X_i)I\{X_i \leq S_{F,m}^{-1}(q_0)\}I\{Y_j \geq S_{G,n}^{-1}(p_0)\}I\{Y_j \leq S_{F,m}^{-1}(q_0)\}$$

$$=\frac{1}{n}\sum_{j=1}^{n}\int_{Y_j}^{S_{F,m}^{-1}(q_0)}dF_m(t)I\{S_{G,n}^{-1}(p_0) \leq Y_j \leq S_{F,m}^{-1}(q_0)\}$$

$$=\frac{1}{n}\sum_{j=1}^{n}[F_m\{S_{F,m}^{-1}(q_0)\} - F_m(Y_j)]I\{S_{G,n}^{-1}(p_0) \leq Y_j \leq S_{F,m}^{-1}(q_0)\}$$

$$=F_m\{S_{F,m}^{-1}(q_0)\}I\{S_{G,n}^{-1}(p_0) \leq Y_j \leq S_{F,m}^{-1}(q_0)\} - \frac{1}{n}\sum_{j=1}^{n}F_m(Y_j)I\{S_{G,n}^{-1}(p_0) \leq Y_j \leq S_{F,m}^{-1}(q_0)\}$$

$$=\int_{G_n^{-1}(1-p_0)}^{F_m^{-1}(1-q_0)}F_m\{F_m^{-1}(1-q_0)\}dG_n(t) - \int_{G_n^{-1}(1-p_0)}^{F_m^{-1}(1-q_0)}F_m(t)dG_n(t)$$

$$=\hat{U}(p_0, q_0). \tag{S.9}$$

On the other hand, we have:

$$U = \int_{S_G\{S_F^{-1}(q_0)\}}^{p_0} S_F\{S_G^{-1}(u)\}du - [p_0 - S_G\{S_F^{-1}(q_0)\}]q_0$$

$$= \int_{S_F^{-1}(q_0)}^{S_G^{-1}(p_0)} S_F(t)dS_G(t) - [p_0 - S_G\{S_F^{-1}(q_0)\}]q_0$$

$$=q_0[S_G\{S_F^{-1}(q_0)\} - p_0] - \int_{S_G^{-1}(p_0)}^{S_F^{-1}(q_0)} S_F(t)dS_G(t)$$

$$=q_0[1 - G\{F^{-1}(1-q_0)\} - p_0] + \int_{G^{-1}(1-p_0)}^{F^{-1}(1-q_0)}\{1 - F(t)\}dG(t)$$

$$=(1-q_0)[G\{F^{-1}(1-q_0)\} - (1-p_0)] - \int_{G^{-1}(1-p_0)}^{F^{-1}(1-q_0)} F(t)dG(t). \tag{S.10}$$

It is obviously that both $\hat{U}$ and $U$ can be composed of two parts. Thus it is natural to write the difference of $\hat{U}$ and $U$ into two parts:

$$\sqrt{m+n}(\hat{U} - U)$$
$$=\sqrt{m+n}\left\{\int_{G_n^{-1}(1-p_0)}^{F_m^{-1}(1-q_0)} F_m\{F_m^{-1}(1-q_0)\}dG_n(t) - (1-q_0)[G\{F^{-1}(1-q_0)\} - (1-p_0)]\right\}$$
$$-\sqrt{m+n}\left\{\int_{G_n^{-1}(1-p_0)}^{F_m^{-1}(1-q_0)} F_m(t)dG_n(t) - \int_{G^{-1}(1-p_0)}^{F^{-1}(1-q_0)} F(t)dG(t)\right\}$$
$$=\sqrt{m+n}\underbrace{\left\{\int_{G_n^{-1}(1-p_0)}^{F_m^{-1}(1-q_0)} F_m\{F_m^{-1}(1-q_0)\}dG_n(t) - \int_{G^{-1}(1-p_0)}^{F^{-1}(1-q_0)} (1-q_0)dG_n(t)\right\}}_{\mathbf{I}}$$
$$-\sqrt{m+n}\underbrace{\left\{\int_{G_n^{-1}(1-p_0)}^{F_m^{-1}(1-q_0)} F_m(t)dG_n(t) - \int_{G^{-1}(1-p_0)}^{F^{-1}(1-q_0)} F(t)dG(t)\right\}}_{\mathbf{II}}. \quad \text{(S.11)}$$

Firstly for term **II** in (S.11), we have

$$\int_{G_n^{-1}(1-p_0)}^{F_m^{-1}(1-q_0)} F_m(t)dG_n(t) - \int_{G^{-1}(1-p_0)}^{F^{-1}(1-q_0)} F(t)dG(t)$$
$$= \underbrace{\int_{-\infty}^{F_m^{-1}(1-q_0)} F_m(t)dG_n(t) - \int_{-\infty}^{F^{-1}(1-q_0)} F(t)dG(t)}_{\mathbf{I_1}}$$
$$- \underbrace{\left\{\int_{-\infty}^{G_n^{-1}(1-p_0)} F_m(t)dG_n(t) - \int_{-\infty}^{G^{-1}(1-p_0)} F(t)dG(t)\right\}}_{\mathbf{II_1}}. \quad \text{(S.12)}$$

For term $\mathbf{I_1}$ in (S.12), according to equation (3.15) in (15),

$$\sqrt{m+n}\left\{\int_{-\infty}^{G_n^{-1}(1-p_0)} F_m(t)dG_n(t) - \int_{-\infty}^{G^{-1}(1-p_0)} F(t)dG(t)\right\}$$
$$=\sqrt{m+n}\int_{-\infty}^{G^{-1}(1-p_0)} F(t)d[G_n(t) - G(t)] + \sqrt{m+n}\int_{-\infty}^{G^{-1}(1-p_0)} [F_m(t) - F(t)]dG(t)$$
$$+ \sqrt{m+n}[1 - p_0 - G_n\{G^{-1}(1-p_0)\}]F\{G^{-1}(1-p_0)\} + o_p(1), m, n \to \infty. \quad \text{(S.13)}$$

For term **II$_1$**, by Theorem 3.7 in (15), we get

$$\sqrt{m+n}\left\{\int_{-\infty}^{F_m^{-1}(1-q_0)}F_m(t)dG_n(t)-\int_{-\infty}^{F^{-1}(1-q_0)}F(t)dG(t)\right\}$$

$$=\sqrt{m+n}\left\{\int_{-\infty}^{F^{-1}(1-q_0)}F(t)dG_n(t)-\int_{-\infty}^{F^{-1}(1-q_0)}F(t)dG(t)\right\}$$

$$+\underbrace{\sqrt{m+n}\left\{\int_{-\infty}^{F_m^{-1}(1-q_0)}F_m(t)dG(t)-\int_{-\infty}^{F^{-1}(1-q_0)}F(t)dG(t)\right\}}_{\mathbf{I_2}}+o_p(1). \tag{S.14}$$

Continuing expanding the term **I$_2$** in (S.14), then we get

$$\sqrt{m+n}\left\{\int_{-\infty}^{F_m^{-1}(1-q_0)}F_m(t)dG(t)-\int_{-\infty}^{F^{-1}(1-q_0)}F(t)dG(t)\right\}$$

$$=\sqrt{m+n}\int_{-\infty}^{F_m^{-1}(1-q_0)}\{F_m(t)-F(t)\}dG(t)$$

$$+\sqrt{m+n}\left\{\int_{-\infty}^{F_m^{-1}(1-q_0)}F(t)dG(t)-\int_{-\infty}^{F^{-1}(1-q_0)}F(t)dG(t)\right\}$$

$$=\underbrace{\sqrt{m+n}\int_{F^{-1}(1-q_0)}^{F_m^{-1}(1-q_0)}\{F_m(t)-F(t)\}dG(t)}_{\mathbf{I_3}}+\sqrt{m+n}\int_{-\infty}^{F^{-1}(1-q_0)}\{F_m(t)-F(t)\}dG(t)$$

$$+\sqrt{m+n}\left\{\int_{-\infty}^{F_m^{-1}(1-q_0)}F(t)dG(t)-\int_{-\infty}^{F^{-1}(1-q_0)}F(t)dG(t)\right\}. \tag{S.15}$$

Combining (S.12)-(S.15), and Lemma S.1.1, we have

$$\int_{G_n^{-1}(1-p_0)}^{F_m^{-1}(1-q_0)} F_m(t)dG_n(t) - \int_{G^{-1}(1-p_0)}^{F^{-1}(1-q_0)} F(t)dG(t)$$

$$=\sqrt{m+n}\int_{-\infty}^{G^{-1}(1-p_0)} F(t)d[G_n(t) - G(t)] + \sqrt{m+n}\int_{-\infty}^{G^{-1}(1-p_0)} [F_m(t) - F(t)]dG(t)$$

$$+ \sqrt{m+n}[1 - p_0 - G_n\{G^{-1}(1-p_0)\}]F\{G^{-1}(1-p_0)\}$$

$$+ \sqrt{m+n}\left\{\int_{-\infty}^{F^{-1}(1-q_0)} F(t)dG_n(t) - \int_{-\infty}^{F^{-1}(1-q_0)} F(t)dG(t)\right\}$$

$$+ \sqrt{m+n}\int_{-\infty}^{F^{-1}(1-q_0)} \{F_m(t) - F(t)\}dG(t)$$

$$+ \sqrt{m+n}\left\{\int_{-\infty}^{F_m^{-1}(1-q_0)} F(t)dG(t) - \int_{-\infty}^{F^{-1}(1-q_0)} F(t)dG(t)\right\} + o_p(1). \qquad \text{(S.16)}$$

As for term **I** of (S.11), we have

$$\int_{G_n^{-1}(1-p_0)}^{F_m^{-1}(1-q_0)} F_m\{F_m^{-1}(1-q_0)\}dG_n(t) - \int_{G^{-1}(1-p_0)}^{F^{-1}(1-q_0)} (1-q_0)dG(t)$$

$$= \underbrace{\int_{-\infty}^{F_m^{-1}(1-q_0)} F_m\{F_m^{-1}(1-q_0)\}dG_n(t) - \int_{-\infty}^{F^{-1}(1-q_0)} (1-q_0)dG(t)}_{\mathbf{I_4}}$$

$$- \underbrace{\left[\int_{-\infty}^{G_n^{-1}(1-p_0)} F_m\{F_m^{-1}(1-q_0)\}dG_n(t) - \int_{-\infty}^{G^{-1}(1-p_0)} (1-q_0)dG(t)\right]}_{\mathbf{II_4}}. \qquad \text{(S.17)}$$

Then we apply Lemma S.1.2 to term $\mathbf{I_4}$ of (S.17), we have

$$\sqrt{m+n}\left[\int_{-\infty}^{F_m^{-1}(1-q_0)} F_m\{F_m^{-1}(1-q_0)\}dG_n(t) - \int_{-\infty}^{F^{-1}(1-q_0)} (1-q_0)dG(t)\right]$$

$$=\sqrt{m+n}\left\{\int_{-\infty}^{F^{-1}(1-q_0)} (1-q_0)dG_n(t) - \int_{-\infty}^{F^{-1}(1-q_0)} (1-q_0)dG(t)\right\}$$

$$+\sqrt{m+n}\left[\int_{-\infty}^{F_m^{-1}(1-q_0)} F_m\{F_m^{-1}(1-q_0)\}dG(t) - \int_{-\infty}^{F^{-1}(1-q_0)} (1-q_0)dG(t)\right] + o_p(1)$$

$$=\sqrt{m+n}\int_{-\infty}^{F^{-1}(1-q_0)} (1-q_0)d\{G_n(t) - G(t)\}$$

$$+\underbrace{\sqrt{m+n}\left[\int_{-\infty}^{F_m^{-1}(1-q_0)} F_m\{F_m^{-1}(1-q_0)\}dG(t) - \int_{-\infty}^{F^{-1}(1-q_0)} (1-q_0)dG(t)\right]}_{\mathbf{I_5}}+o_p(1). \qquad (\text{S.18})$$

Similarly, for term $\mathbf{I_5}$ of (S.18), we have

$$\sqrt{m+n}\left[\int_{-\infty}^{F_m^{-1}(1-q_0)} F_m\{F_m^{-1}(1-q_0)\}dG(t) - \int_{-\infty}^{F^{-1}(1-q_0)} (1-q_0)dG(t)\right]$$

$$=\sqrt{m+n}\int_{-\infty}^{F_m^{-1}(1-q_0)} [F_m\{F_m^{-1}(1-q_0)\} - (1-q_0)]dG(t)$$

$$+\sqrt{m+n}\left\{\int_{-\infty}^{F_m^{-1}(1-q_0)} (1-q_0)dG(t) - \int_{-\infty}^{F^{-1}(1-q_0)} (1-q_0)dG(t)\right\}$$

$$=\underbrace{\sqrt{m+n}\int_{F^{-1}(1-q_0)}^{F_m^{-1}(1-q_0)} [F_m\{F_m^{-1}(1-q_0)\} - (1-q_0)]dG(t)}_{\mathbf{I_6}}$$

$$+\sqrt{m+n}\int_{-\infty}^{F^{-1}(1-q_0)} [F_m\{F_m^{-1}(1-q_0)\} - (1-q_0)]dG(t)$$

$$+\sqrt{m+n}\left\{\int_{-\infty}^{F_m^{-1}(1-q_0)} (1-q_0)dG(t) - \int_{-\infty}^{F^{-1}(1-q_0)} (1-q_0)dG(t)\right\}. \qquad (\text{S.19})$$

combining equations (S.18), (S.19) and Lemma S.1.3, we get

$$\int_{-\infty}^{F_m^{-1}(1-q_0)} F_m\{F_m^{-1}(1-q_0)\}dG_n(t) - \int_{-\infty}^{F^{-1}(1-q_0)} (1-q_0)dG(t)$$

$$=\sqrt{m+n}\int_{-\infty}^{F^{-1}(1-q_0)} (1-q_0)d\{G_n(t) - G(t)\}$$

$$+ \sqrt{m+n}\int_{-\infty}^{F^{-1}(1-q_0)} [F_m\{F_m^{-1}(1-q_0)\} - (1-q_0)]dG(t).$$

$$+ \sqrt{m+n}\left\{\int_{-\infty}^{F_m^{-1}(1-q_0)} (1-q_0)dG(t) - \int_{-\infty}^{F^{-1}(1-q_0)} (1-q_0)dG(t)\right\} \quad \text{(S.20)}$$

Then we apply the same process as (S.18), (S.19) and Lemma S.1.3 to term $\mathbf{II_4}$ of equation (S.17), we obtain

$$\int_{-\infty}^{G_n^{-1}(1-p_0)} F_m\{F_m^{-1}(1-q_0)\}dG_n(t) - \int_{-\infty}^{G^{-1}(1-p_0)} (1-q_0)dG(t)$$

$$=\sqrt{m+n}\int_{-\infty}^{G^{-1}(1-p_0)} (1-p_0)d[G_n(t) - G(t)] + \sqrt{m+n}\int_{-\infty}^{G^{-1}(1-p_0)} F_m\{F_m^{-1}(1-q_0)\}dG(t)$$

$$+ \sqrt{m+n}\left[(1-q_0)G\{G_n^{-1}(1-p_0)\} - 2(1-q_0)(1-p_0)\right]. \quad \text{(S.21)}$$

Therefore, with the result of (S.20) and (S.21), (S.17) becomes

$$\int_{G_n^{-1}(1-p_0)}^{F_m^{-1}(1-q_0)} F_m\{F_m^{-1}(1-q_0)\}dG_n(t) - \int_{G^{-1}(1-p_0)}^{F^{-1}(1-q_0)} (1-q_0)dG_n(t)$$

$$=\sqrt{m+n}\int_{-\infty}^{F^{-1}(1-q_0)} (1-q_0)d\{G_n(t) - G(t)\}$$

$$+ \sqrt{m+n}\int_{-\infty}^{F^{-1}(1-q_0)} [F_m\{F_m^{-1}(1-q_0)\} - (1-q_0)]dG(t)$$

$$+ \sqrt{m+n}\left\{\int_{-\infty}^{F_m^{-1}(1-q_0)} (1-q_0)dG(t) - \int_{-\infty}^{F^{-1}(1-q_0)} (1-q_0)dG(t)\right\} \quad \text{(S.22)}$$

$$- (\sqrt{m+n}\int_{-\infty}^{G^{-1}(1-p_0)} (1-p_0)d[G_n(t) - G(t)] + \sqrt{m+n}\int_{-\infty}^{G^{-1}(1-p_0)} F_m\{F_m^{-1}(1-q_0)\}dG(t)$$

$$+ \sqrt{m+n}\left[(1-q_0)G\{G_n^{-1}(1-p_0)\} - 2(1-q_0)(1-p_0)\right]). \quad \text{(S.23)}$$

Thus, above all, (S.12) turns to

$$
\begin{aligned}
&\sqrt{m+n}(\hat{U} - U) \\
=&\sqrt{m+n} \int_{-\infty}^{F^{-1}(1-q_0)} (1-q_0) d\{G_n(t) - G(t)\} \\
&+ \sqrt{m+n} \int_{-\infty}^{F^{-1}(1-q_0)} [F_m\{F_m^{-1}(1-q_0)\} - (1-q_0)] dG(t) \\
&+ \underbrace{\sqrt{m+n} \left\{ \int_{-\infty}^{F_m^{-1}(1-q_0)} (1-q_0) dG(t) - \int_{-\infty}^{F^{-1}(1-q_0)} (1-q_0) dG(t) \right\}}_{I_7} \\
&- (\sqrt{m+n} \int_{-\infty}^{G^{-1}(1-p_0)} (1-q_0) d[G_n(t) - G(t)] \\
&+ \sqrt{m+n} \int_{-\infty}^{G^{-1}(1-p_0)} [F_m\{F_m^{-1}(1-q_0)\} - (1-q_0)] dG(t) \\
&+ (1-q_0) \underbrace{\sqrt{m+n} \left[ G\{G_n^{-1}(1-p_0)\} - (1-p_0) \right]}_{II_7} ) \\
&+ \sqrt{m+n} \int_{-\infty}^{G^{-1}(1-p_0)} F(t) d[G_n(t) - G(t)] + \sqrt{m+n} \int_{-\infty}^{G^{-1}(1-p_0)} [F_m(t) - F(t)] dG(t) \\
&+ \sqrt{m+n}[1 - p_0 - G_n\{G^{-1}(1-p_0)\}] F\{G^{-1}(1-p_0)\} \\
&- \sqrt{m+n} \left\{ \int_{-\infty}^{F^{-1}(1-q_0)} F(t) dG_n(t) - \int_{-\infty}^{F^{-1}(1-q_0)} F(t) dG(t) \right\} \\
&- \sqrt{m+n} \int_{-\infty}^{F^{-1}(1-q_0)} \{F_m(t) - F(t)\} dG(t) \\
&- \underbrace{\sqrt{m+n} \left\{ \int_{-\infty}^{F_m^{-1}(1-q_0)} F(t) dG(t) - \int_{-\infty}^{F^{-1}(1-q_0)} F(t) dG(t) \right\}}_{III_7}. \quad\quad\quad \text{(S.24)}
\end{aligned}
$$

On the other hand, with similar proof procedures to get Lemma S.1.2, we can reach the same conclusion,

$$\sqrt{m+n}(P_m f_m - P f_0) = \sqrt{m+n}(P_m f_0 - P f_0) + \sqrt{m+n}(P f_m - P f_0) + o_p(1), \quad m, n \to \infty. \quad \text{(S.25)}$$

Then we apply (S.25) to term $\sqrt{m+n}[F_m\{F_m^{-1}(1-q_0)\} - (1-q_0)]$,

$$\sqrt{m+n}[F_m\{F_m^{-1}(1-q_0)\} - (1-q_0)]$$
$$=\sqrt{m+n}\left\{\int_{-\infty}^{F_m^{-1}(1-q_0)} dF_m(t) - \int_{-\infty}^{F^{-1}(1-q_0)} dF(t)\right\}$$
$$=\sqrt{m+n}\left\{\int_{-\infty}^{F^{-1}(1-q_0)} dF_m(t) - \int_{-\infty}^{F^{-1}(1-q_0)} dF(t)\right\}$$
$$+ \sqrt{m+n}\left\{\int_{-\infty}^{F^{-1}(1-q_0)} dF_m(t) - \int_{-\infty}^{F^{-1}(1-q_0)} dF(t)\right\} + o_p(1)$$
$$=\sqrt{m+n}[F_m\{F^{-1}(1-q_0)\} - (1-q_0)] + \sqrt{m+n}[F\{F_m^{-1}(1-q_0)\} - (1-q_0)] + o_p(1). \quad \text{(S.26)}$$

Equivalently, it means

$$\sqrt{m+n}F_m\{F_m^{-1}(1-q_0)\}$$
$$=\sqrt{m+n}(1-q_0) + \sqrt{m+n}[F_m\{F^{-1}(1-q_0)\} - (1-q_0)]$$
$$+ \underbrace{\sqrt{m+n}[F\{F_m^{-1}(1-q_0)\} - (1-q_0)]}_{\mathbf{I_8}} + o_p(1). \quad \text{(S.27)}$$

Moreover, we can apply the proof methods of Lemma S.1.4 and Lemma S.1.5 to term $\mathbf{II_7}$ of (S.24) and term $\mathbf{I_8}$ of (S.27), we have

$$\sqrt{m+n}[F\{F_m^{-1}(1-q_0)\} - (1-q_0)]$$
$$=\sqrt{m+n}\left\{\int_{-\infty}^{F_m^{-1}(1-q_0)} dF(t) - \int_{-\infty}^{F^{-1}(1-q_0)} dF(t)\right\}$$
$$=\sqrt{m+n}[1 - q_0 - F_m\{F^{-1}(1-q_0)\}] + o_p(1), \quad \text{(S.28)}$$

and

$$\sqrt{m+n}[G\{G_n^{-1}(1-p_0)\} - (1-p_0)]$$
$$=\sqrt{m+n}\left\{\int_{-\infty}^{G_n^{-1}(1-p_0)} dG(t) - \int_{-\infty}^{G^{-1}(1-p_0)} dG(t)\right\}$$
$$=\sqrt{m+n}[1 - p_0 - G_n\{G^{-1}(1-p_0)\}] + o_p(1). \tag{S.29}$$

Therefore, with (S.24)-(S.29), (S.24) becomes

$$\sqrt{m+n}(\hat{U} - U)$$
$$= \int_{F^{-1}(1-q_0)}^{G^{-1}(1-p_0)} F_m(t)dG(t) - \int_{F^{-1}(1-q_0)}^{G^{-1}(1-p_0)} F(t)dG(t)$$
$$+ (1-q_0)G_n\{F^{-1}(1-q_0)\} + \int_{-\infty}^{G^{-1}(1-p_0)} F(t)dG_n(t) - G_n\{G^{-1}(1-p_0)\}]F\{G^{-1}(1-p_0)\}$$
$$- \int_{-\infty}^{F^{-1}(1-q_0)} F(t)dG_n(t)$$
$$- [(1-q_0)G\{F^{-1}(1-q_0)\} + \int_{-\infty}^{G^{-1}(1-p_0)} F(t)dG(t) - (1-p_0)F\{G^{-1}(1-p_0)\}$$
$$- \int_{-\infty}^{F^{-1}(1-q_0)} F(t)dG(t)] + o_p(1). \tag{S.30}$$

Next, through integration by parts, we have

$$\int_{-\infty}^{G^{-1}(1-p_0)} F(t)dG_n(t) - G_n\{G^{-1}(1-p_0)\}]F\{G^{-1}(1-p_0)\}$$
$$= \int_{-\infty}^{G^{-1}(1-p_0)} G_n(t)dF(t), \tag{S.31}$$

$$\int_{-\infty}^{G^{-1}(1-p_0)} F(t)dG(t) - (1-p_0)F\{G^{-1}(1-p_0)\}$$
$$= \int_{-\infty}^{G^{-1}(1-p_0)} G(t)dF(t), \tag{S.32}$$

$$(1-q_0)G_n\{F^{-1}(1-q_0)\} - \int_{-\infty}^{F^{-1}(1-q_0)} F(t)dG_n(t)$$
$$= -\int_{-\infty}^{F^{-1}(1-q_0)} G_n(t)dF(t), \tag{S.33}$$

and

$$(1-q_0)G\{F^{-1}(1-q_0)\} - \int_{-\infty}^{F^{-1}(1-q_0)} F(t)dG(t)$$
$$= -\int_{-\infty}^{F^{-1}(1-q_0)} G(t)dF(t). \tag{S.34}$$

Then combining (S.30)-(S.34), we get

$$\sqrt{m+n}(\hat{U} - U)$$
$$= \int_{F^{-1}(1-q_0)}^{G^{-1}(1-p_0)} F_m(t)dG(t) - \int_{F^{-1}(1-q_0)}^{G^{-1}(1-p_0)} F(t)dG(t)$$
$$+ \int_{F^{-1}(1-q_0)}^{G^{-1}(1-p_0)} G_n(t)dF(t) - \int_{F^{-1}(1-q_0)}^{G^{-1}(1-p_0)} G(t)dF(t) + o_p(1). \tag{S.35}$$

For now we can write (S.35) into $\sqrt{m+n}(T_m - \mu_1 + T_n - \mu_2) + o_p(1)$, where

$$T_m = \int_{F^{-1}(1-q_0)}^{G^{-1}(1-p_0)} F_m(t)dG(t),$$

$$\mu_1 = \int_{F^{-1}(1-q_0)}^{G^{-1}(1-p_0)} F(t)dG(t),$$

$$T_n = \int_{F^{-1}(1-q_0)}^{G^{-1}(1-p_0)} G_n(t)dF(t),$$

and

$$\mu_2 = \int_{F^{-1}(1-q_0)}^{G^{-1}(1-p_0)} G(t)dF(t).$$

Consider $T_m$ first, by rewriting, we have,

$$
\begin{aligned}
T_m &= \int_{F^{-1}(1-q_0)}^{G^{-1}(1-p_0)} F_m(t) dG(t) \\
&= \int_{F^{-1}(1-q_0)}^{G^{-1}(1-p_0)} \frac{1}{m} \sum_{i=1}^{m} I(X_i < t) dG(t) \\
&= \frac{1}{m} \sum_{i=1}^{m} \int_{F^{-1}(1-q_0)}^{G^{-1}(1-p_0)} I(X_i < t) dG(t).
\end{aligned}
\tag{S.36}
$$

Note that $T_m$ is in fact a sum of i.i.d. random variable, thus

$$
\begin{aligned}
E(T_m) &= E\left\{ \int_{F^{-1}(1-q_0)}^{G^{-1}(1-p_0)} I(X < t) dG(t) \right\} \\
&= \int_{-\infty}^{\infty} \left\{ \int_{F^{-1}(1-q_0)}^{G^{-1}(1-p_0)} I(X < t) dG(t) \right\} dF(X) \\
&= \int_{-\infty}^{\infty} \left\{ \int_{-\infty}^{\infty} I\{F^{-1}(1-q_0) \leq t \leq G^{-1}(1-p_0)\} I(X < t) dG(t) \right\} dF(X) \\
&= \int_{-\infty}^{\infty} I\{F^{-1}(1-q_0) \leq t \leq G^{-1}(1-p_0)\} \left\{ \int_{-\infty}^{\infty} I(X < t) dF(X) \right\} dG(t) \\
&= \int_{F^{-1}(1-q_0)}^{G^{-1}(1-p_0)} F(t) dG(t) \\
&= \mu_1.
\end{aligned}
\tag{S.37}
$$

Denote $\sigma_3^2$ to be the variance of $T_m$, then

$$\sigma_3^2 = Var\left\{\int_{F^{-1}(1-q_0)}^{G^{-1}(1-p_0)} I(X < t)dG(t)\right\}$$

$$= E\left\{\int_{G^{-1}(1-p_0)}^{F^{-1}(1-q_0)} I(X < t)dG(t)\right\}^2 - \left[E\left\{\int_{G^{-1}(1-p_0)}^{F^{-1}(1-q_0)} I(X < t)dG(t)\right\}\right]^2$$

$$= \int_{-\infty}^{\infty}\left\{\int_{G^{-1}(1-p_0)}^{F^{-1}(1-q_0)} I(X < t)dG(t)\right\}^2 dF(X) - \left\{\int_{G^{-1}(1-p_0)}^{F^{-1}(1-q_0)} F(t)dG(t)\right\}^2$$

$$= \int_{-\infty}^{\infty}\left\{\int_{-\infty}^{\infty} I\{G^{-1}(1-p_0) \le t \le F^{-1}(1-q_0)\}I(X < t)dG(t)\right\}^2 dF(X)$$

$$- \left\{\int_{G^{-1}(1-p_0)}^{F^{-1}(1-q_0)} F(t)dG(t)\right\}^2$$

$$= \int_{-\infty}^{\infty} [\int_{-\infty}^{\infty} I\{X \le G^{-1}(1-p_0)\}I\{G^{-1}(1-p_0) \le t \le F^{-1}(1-q_0)\}dG(t)+$$

$$+ \int_{-\infty}^{\infty} I\{F^{-1}(1-q_0) > X > G^{-1}(1-p_0)\}I\{X \le t \le F^{-1}(1-q_0)\}dG(t)]^2 dF(X)$$

$$- \left\{\int_{G^{-1}(1-p_0)}^{F^{-1}(1-q_0)} F(t)dG(t)\right\}^2$$

$$= \int_{-\infty}^{\infty}(I\{X \le G^{-1}(1-p_0)\}[G\{F^{-1}(1-q_0)\} - (1-p_0)]+$$

$$+ I\{F^{-1}(1-q_0) > X > G^{-1}(1-p_0)\}\{G\{F^{-1}(1-q_0)\} - G(X)\})^2 dF(X)$$

$$- \left\{\int_{G^{-1}(1-p_0)}^{F^{-1}(1-q_0)} F(t)dG(t)\right\}^2$$

$$= \int_{-\infty}^{\infty}(I\{X \le G^{-1}(1-p_0)\}[G\{F^{-1}(1-q_0)\} - (1-p_0)\}]^2+$$

$$+ I\{F^{-1}(1-q_0) > X > G^{-1}(1-p_0)\}[G\{F^{-1}(1-q_0)\} - G(X)]^2)dF(X)$$

$$- \left\{\int_{G^{-1}(1-p_0)}^{F^{-1}(1-q_0)} F(t)dG(t)\right\}^2$$

$$= \int_{-\infty}^{\infty}(I\{t \le G^{-1}(1-p_0)\}[G\{F^{-1}(1-q_0)\} - (1-p_0)]^2+$$

$$+ I\{F^{-1}(1-q_0) > t > G^{-1}(1-p_0)\}[G\{F^{-1}(1-q_0)\} - G(t)]^2)dF(t)$$

$$- \left\{\int_{G^{-1}(1-p_0)}^{F^{-1}(1-q_0)} F(t)dG(t)\right\}^2$$

$$= F\{G^{-1}(1-p_0)\}[G\{F^{-1}(1-q_0)\} - (1-p_0)]^2+$$

$$+ \int_{G^{-1}(1-p_0)}^{F^{-1}(1-q_0)} [G\{F^{-1}(1-q_0)\} - G(t)]^2 dF(t)$$

$$- \left\{\int_{G^{-1}(1-p_0)}^{F^{-1}(1-q_0)} F(t)dG(t)\right\}^2.$$

Then by Lindeberg-Levy Central Limit Theorem, we have

$$\sqrt{m}(T_m - \mu_1) \xrightarrow{d} N(0, \sigma_3^2), \quad m, n \to \infty.$$

Let m, n be sequences of integers such that $\frac{m}{m+n} \to \lambda$, further note that

$$\sqrt{\frac{n}{m}} = \sqrt{\frac{m+n}{m} - 1} \to \sqrt{\frac{1}{\lambda} - 1}, \quad m, n \to \infty.$$

Thus, by Slutsky's Theorem, we have,

$$\sqrt{n}(T_m - \mu_1) \xrightarrow{d} N\left\{0, \left(\frac{1}{\lambda} - 1\right)\sigma_3^2\right\}, \quad m, n \to \infty. \tag{S.38}$$

Then consider $T_n$,

$$\begin{aligned} T_n &= \int_{F^{-1}(1-q_0)}^{G^{-1}(1-p_0)} G_n(t) dF(t) \\ &= \int_{F^{-1}(1-q_0)}^{G^{-1}(1-p_0)} \frac{1}{n} \sum_{j=1}^{n} I(Y_j \le t) dF(t) \\ &= \frac{1}{n} \sum_{j=1}^{n} \int_{F^{-1}(1-q_0)}^{G^{-1}(1-p_0)} I(Y_j \le t) dF(t). \end{aligned} \tag{S.39}$$

Since $T_n$ is also a sum of i.i.d. random variables, then

$$\begin{aligned} E(T_n) &= E\left\{\int_{F^{-1}(1-q_0)}^{G^{-1}(1-p_0)} I(Y \le t) dF(t)\right\} \\ &= \int_{-\infty}^{\infty} \left\{\int_{F^{-1}(1-q_0)}^{G^{-1}(1-p_0)} I(Y \le t) dF(t)\right\} dG(Y) \\ &= \int_{-\infty}^{\infty} \left[\int_{-\infty}^{\infty} I\{F^{-1}(1-q_0) \le t \le G^{-1}(1-p_0)\} I(Y \le t) dF(t)\right] dG(Y) \\ &= \int_{-\infty}^{\infty} \int_{-\infty}^{\infty} I\{Y \le t\} dG(Y) I\{F^{-1}(1-q_0) \le t \le G^{-1}(1-p_0)\} dF(t) \\ &= \int_{F^{-1}(1-q_0)}^{G^{-1}(1-p_0)} G(t) dF(t). \end{aligned} \tag{S.40}$$

Denote $\sigma_4^2$ to be the variance of $T_n$, then

$$\sigma_4^2 = Var\left\{\int_{F^{-1}(1-q_0)}^{G^{-1}(1-p_0)} I(Y<t)dF(t)\right\}$$

$$= E\left\{\int_{G^{-1}(1-p_0)}^{F^{-1}(1-q_0)} I(Y<t)dF(t)\right\}^2 - \left[E\left\{\int_{G^{-1}(1-p_0)}^{F^{-1}(1-q_0)} I(Y<t)dF(t)\right\}\right]^2$$

$$= \int_{-\infty}^{\infty}\left\{\int_{G^{-1}(1-p_0)}^{F^{-1}(1-q_0)} I(Y<t)dF(t)\right\}^2 dG(Y) - \left\{\int_{G^{-1}(1-p_0)}^{F^{-1}(1-q_0)} G(t)dF(t)\right\}^2$$

$$= \int_{-\infty}^{\infty}\left\{\int_{-\infty}^{\infty} I\{G^{-1}(1-p_0) \le t \le F^{-1}(1-q_0)\}I(Y<t)dF(t)\right\}^2 dG(Y)$$

$$- \left\{\int_{G^{-1}(1-p_0)}^{F^{-1}(1-q_0)} G(t)dF(t)\right\}^2$$

$$= \int_{-\infty}^{\infty}\{\int_{-\infty}^{\infty} I\{G^{-1}(1-p_0) \le t \le F^{-1}(1-q_0)\}I\{Y \le G^{-1}(1-p_0)\}dF(t)$$

$$+ \int_{-\infty}^{\infty} I\{Y \le t \le F^{-1}(1-q_0)\}I\{G^{-1}(1-p_0) < Y < F^{-1}(1-q_0)\}dF(t)\}^2 dG(Y)$$

$$- \left\{\int_{G^{-1}(1-p_0)}^{F^{-1}(1-q_0)} G(t)dF(t)\right\}^2$$

$$= \int_{-\infty}^{\infty}\{[1-q_0-F\{G^{-1}(1-p_0)\}]I\{Y \le G^{-1}(1-p_0)\}$$

$$+ \{1-q_0-F(Y)\}I\{G^{-1}(1-p_0) \le Y \le F^{-1}(1-q_0)\}\}^2 dG(Y)$$

$$- \left\{\int_{G^{-1}(1-p_0)}^{F^{-1}(1-q_0)} G(t)dF(t)\right\}^2$$

$$= \int_{-\infty}^{\infty}[1-q_0-F\{G^{-1}(1-p_0)\}]^2 I\{Y \le G^{-1}(1-p_0)\}$$

$$+ \{1-q_0-F(Y)\}^2 I\{G^{-1}(1-p_0) \le Y \le F^{-1}(1-q_0)\})dG(Y)$$

$$- \left\{\int_{G^{-1}(1-p_0)}^{F^{-1}(1-q_0)} G(t)dF(t)\right\}^2$$

$$= \int_{-\infty}^{\infty}[1-q_0-F\{G^{-1}(1-p_0)\}]^2 I\{t \le G^{-1}(1-p_0)\}$$

$$+ \{1-q_0-F(t)\}^2 I\{G^{-1}(1-p_0) \le t \le F^{-1}(1-q_0)\})dG(t)$$

$$- \left\{\int_{G^{-1}(1-p_0)}^{F^{-1}(1-q_0)} G(t)dF(t)\right\}^2$$

$$= [1-q_0-F\{G^{-1}(1-p_0)\}]^2(1-p_0)$$

$$+ \int_{G^{-1}(1-p_0)}^{F^{-1}(1-q_0)} \{1-q_0-F(t)\}^2 dG(t)$$

$$- \left\{\int_{G^{-1}(1-p_0)}^{F^{-1}(1-q_0)} G(t)dF(t)\right\}^2.$$

Thus by Lindeberg-Levy Central Limit Theorem, we have

$$\sqrt{n}(T_n - \mu_2) \xrightarrow{d} N(0, \sigma_4^2), \quad m, n \to \infty. \tag{S.41}$$

Because $T_m$ and $T_n$ are independent, and $\hat{U} - U = T_m - \mu_1 + T_n - \mu_2$, thus

$$\sqrt{n}(\hat{U} - U) \xrightarrow{d} N\left\{0, \left(\frac{1}{\lambda} - 1\right)\sigma_3^2 + \sigma_4^2\right\}, \quad m, n \to \infty. \tag{S.42}$$

Since

$$\sqrt{\frac{m+n}{n}} = \frac{1}{\sqrt{\frac{n}{m+n}}} \to \sqrt{\frac{1}{1-\lambda}}, 0 < \lambda < 1, m, n \to \infty.$$

Then by Slutsky's Theorem,

$$\sqrt{m+n}(\hat{U} - U) \xrightarrow{d} N\left\{0, \frac{\sigma_3^2}{\lambda} + \frac{\sigma_4^2}{1-\lambda}\right\}, \quad m, n \to \infty, \tag{S.43}$$

where

$$\sigma_3^2 = F\{G^{-1}(1-p_0)\}[G\{F^{-1}(1-q_0)\} - (1-p_0)]^2 +$$
$$+ \int_{G^{-1}(1-p_0)}^{F^{-1}(1-q_0)} [G\{F^{-1}(1-q_0)\} - G(t)]^2 dF(t)$$
$$- \left\{\int_{G^{-1}(1-p_0)}^{F^{-1}(1-q_0)} F(t) dG(t)\right\}^2,$$

and

$$\sigma_4^2 = [1 - q_0 - F\{G^{-1}(1-p_0)\}]^2 (1-p_0)$$
$$+ \int_{G^{-1}(1-p_0)}^{F^{-1}(1-q_0)} \{1 - q_0 - F(t)\}^2 dG(t)$$
$$- \left\{\int_{G^{-1}(1-p_0)}^{F^{-1}(1-q_0)} G(t) dF(t)\right\}^2.$$

Then we finished the proof Theorem 4.1.

## S.3  Proof of Theorem 4.2

Firstly, for all those ROC area indexes $\hat{\theta}^*$ we referred to in this paper, we know $\sqrt{n+m}(\hat{\theta}^* - \theta)$ converges to a normal random variable in distribution. Secondly, according to equation (6.7) in Page 47 of (14), we have

$$v_{boot}^2 \to \frac{\sigma_3^2}{\lambda} + \frac{\sigma_4^2}{1-\lambda}, \quad B \to \infty.$$

Therefore we can then prove Theorem 4.2 directly by using Slutsky's Theorem.

## S.4  Proof of Theorem 4.3

Our proof strategy is to apply Theorem 2 in (16), which establishes sufficient conditions for uniqueness and consistency for solution to likelihood estimation. To achieve this, we need to prove the four conditions in (16) are satisfied. We briefly introduce these conditions in our notation as follows,

(**F1**) $\partial S_{m,n}(\boldsymbol{\beta})/\partial \boldsymbol{\beta}$ exists and is continuous in $N_\delta(\boldsymbol{\beta}_0)$.

(**F2**) $\partial S_{m,n}(\boldsymbol{\beta})/\partial \boldsymbol{\beta} \to_p \mathbb{E}\{\partial S_{i,j}(\boldsymbol{\beta})/\partial \boldsymbol{\beta}\}$ uniformly in $N_\delta(\boldsymbol{\beta}_0)$, as $m, n \to \infty$.

(**F3**) With probability tends to one, as $m, n \to \infty$, $\partial S_{m,n}(\boldsymbol{\beta}_0)/\partial \boldsymbol{\beta}$ is negative definite.

(**F4**) $\mathbb{E} S_{m,n}(\boldsymbol{\beta}_0) = 0$.

We first prove Condition F3 is satisfied. From triangle inequality, we have, for any $\epsilon > 0$,

$$\begin{aligned}
& P\{|\partial S_{m,n}(\boldsymbol{\beta})/\partial \boldsymbol{\beta} - \mathbb{E}\{\partial S_{i,j}(\boldsymbol{\beta})/\partial \boldsymbol{\beta}\}| > \epsilon\} \\
\leq & P\{|\partial S_{m,n}(\boldsymbol{\beta})/\partial \boldsymbol{\beta} - \frac{1}{m}\sum_{i=1}^m \mathbb{E}\{\partial S_{i,j}(\boldsymbol{\beta})/\partial \boldsymbol{\beta}|X_i\}| > \epsilon/2\} \\
& + P\{|\frac{1}{m}\sum_{i=1}^m \mathbb{E}\{\partial S_{i,j}(\boldsymbol{\beta})/\partial \boldsymbol{\beta}|X_i\} - \mathbb{E}\{\partial S_{i,j}(\boldsymbol{\beta})/\partial \boldsymbol{\beta}\}| > \epsilon/2\}
\end{aligned} \tag{S.44}$$

where $\mathbb{E}\{\partial S_{i,j}(\boldsymbol{\beta})/\partial \boldsymbol{\beta}|X_i\}$ is independent across $i \in \{1,\ldots,m\}$ and random in terms of $X_i$. For the first term in (S.44), we get

$$\mathbb{E}\left|\partial S_{m,n}(\boldsymbol{\beta})/\partial \boldsymbol{\beta} - \frac{1}{m}\sum_{i=1}^{m}\mathbb{E}\{\partial S_{i,j}(\boldsymbol{\beta})/\partial \boldsymbol{\beta}|X_i\}\right|$$

$$=\mathbb{E}\left|\frac{1}{m}\sum_{i=1}^{m}\frac{1}{n}\sum_{j=1}^{n}\partial S_{i,j}(\boldsymbol{\beta})/\partial \boldsymbol{\beta} - \frac{1}{m}\sum_{i=1}^{m}\mathbb{E}\{\partial S_{i,j}(\boldsymbol{\beta})/\partial \boldsymbol{\beta}|X_i\}\right|$$

$$\leq \frac{1}{m}\sum_{i=1}^{m}E\left|\frac{1}{n}\sum_{j=1}^{n}\partial S_{i,j}(\boldsymbol{\beta})/\partial \boldsymbol{\beta} - \mathbb{E}\{\partial S_{i,j}(\boldsymbol{\beta})/\partial \boldsymbol{\beta}|X_i\}\right|$$

where the inequality is obtained from $\partial S_{i,j}(\boldsymbol{\beta})/\partial \boldsymbol{\beta}$ are i.i.d. across all $j$ given $i$. Therefore, we have

$$P\{|\partial S_{m,n}(\boldsymbol{\beta})/\partial \boldsymbol{\beta} - \frac{1}{m}\sum_{i=1}^{m}\mathbb{E}\{\partial S_{i,j}(\boldsymbol{\beta})/\partial \boldsymbol{\beta}|X_i\}| > \epsilon/2\} \to 0, \text{ as } m,n \to \infty, \quad \text{(S.45)}$$

by weak law of large numbers and convergence in probability is weaker than convergence in mean. With similar arguments, we obtain

$$P\{|\frac{1}{m}\sum_{i=1}^{m}\mathbb{E}\{\partial S_{i,j}(\boldsymbol{\beta})/\partial \boldsymbol{\beta}|X_i\} - \mathbb{E}\{\partial S_{i,j}(\boldsymbol{\beta})/\partial \boldsymbol{\beta}\}| > \epsilon/2\} \to 0, \text{ as } m,n \to \infty. \quad \text{(S.46)}$$

Combine (S.45) and (S.46), we get

$$P\{|\partial S_{m,n}(\boldsymbol{\beta})/\partial \boldsymbol{\beta} - \mathbb{E}\{\partial S_{i,j}(\boldsymbol{\beta})/\partial \boldsymbol{\beta}\}| > \epsilon\} \to 0, \quad \text{(S.47)}$$

as $m,n \to \infty$. Together with Assumption 4, the proof to Condition F3 is complete.

Next, we turn to Condition F2. There exists a union of finite balls with known radius that cover $N_\delta(\boldsymbol{\beta}_0)$. Define balls as $\odot_k$ for all $k \in \{1,\ldots,K\}$ with center $\boldsymbol{\beta}_k$ and radius less than $r$. The finite cover of $N_\delta(\boldsymbol{\beta}_0)$ is $\bigcup_{k=1}^{K} \odot_k$. By triangle inequality, we have

$$\sup_{\boldsymbol{\beta} \in N_\delta(\boldsymbol{\beta}_0)} \left| \partial S_{m,n}(\boldsymbol{\beta})/\partial\boldsymbol{\beta} - \mathbb{E}\{\partial S_{i,j}(\boldsymbol{\beta})/\partial\boldsymbol{\beta}\} \right|$$

$$= \max_k \sup_{\boldsymbol{\beta} \in \odot_k} \Big| \partial S_{m,n}(\boldsymbol{\beta})/\partial\boldsymbol{\beta} - \partial S_{m,n}(\boldsymbol{\beta}_k)/\partial\boldsymbol{\beta}$$

$$+ \mathbb{E}\{\partial S_{i,j}(\boldsymbol{\beta}_k)/\partial\boldsymbol{\beta}\} - \mathbb{E}\{\partial S_{i,j}(\boldsymbol{\beta})/\partial\boldsymbol{\beta}\}$$

$$+ \partial S_{m,n}(\boldsymbol{\beta}_k)/\partial\boldsymbol{\beta} - \mathbb{E}\{\partial S_{i,j}(\boldsymbol{\beta}_k)/\partial\boldsymbol{\beta}\} \Big|$$

$$\leq \max_k \sup_{\boldsymbol{\beta} \in \odot_k} \left| \partial S_{m,n}(\boldsymbol{\beta})/\partial\boldsymbol{\beta} - \partial S_{m,n}(\boldsymbol{\beta}_k)/\partial\boldsymbol{\beta} \right|$$

$$+ \max_k \sup_{\boldsymbol{\beta} \in \odot_k} \left| \mathbb{E}\{\partial S_{i,j}(\boldsymbol{\beta}_k)/\partial\boldsymbol{\beta}\} - \mathbb{E}\{\partial S_{i,j}(\boldsymbol{\beta})/\partial\boldsymbol{\beta}\} \right|$$

$$+ \max_k \left| \partial S_{m,n}(\boldsymbol{\beta}_k)/\partial\boldsymbol{\beta} - \mathbb{E}\{\partial S_{i,j}(\boldsymbol{\beta}_k)/\partial\boldsymbol{\beta}\} \right| \tag{S.48}$$

For the last term in (S.48), we have, for $\epsilon > 0$ and $\tau > 0$,

$$P\left\{ \max_k \left| \partial S_{m,n}(\boldsymbol{\beta}_k)/\partial\boldsymbol{\beta} - \mathbb{E}\{\partial S_{i,j}(\boldsymbol{\beta}_k)/\partial\boldsymbol{\beta}\} \right| > \epsilon/2 \right\}$$

$$\leq \sum_{k=1}^{K} P\left\{ \left| \partial S_{m,n}(\boldsymbol{\beta}_k)/\partial\boldsymbol{\beta} - \mathbb{E}\{\partial S_{i,j}(\boldsymbol{\beta}_k)/\partial\boldsymbol{\beta}\} \right| > \epsilon/2 \right\}$$

$$< \sum_{k=1}^{K} \tau/K = \tau, \tag{S.49}$$

where the second inequality is obtained from (S.47). For the first term in (S.48), by mean value theorem, we have

$$\max_k \sup_{\boldsymbol{\beta} \in \odot_k} \left| \partial S_{m,n}(\boldsymbol{\beta})/\partial\boldsymbol{\beta} - \partial S_{m,n}(\boldsymbol{\beta}_k)/\partial\boldsymbol{\beta} \right|$$

$$= \max_k \sup_{\boldsymbol{\beta} \in \odot_k} (\boldsymbol{\beta} - \boldsymbol{\beta}_k) \frac{\partial}{\partial\boldsymbol{\beta}} \frac{\partial S_{m,n}(\boldsymbol{\beta}^*)}{\partial\boldsymbol{\beta}}$$

$$\leq rM_1, \tag{S.50}$$

for certain $\beta^* \in \odot_k$. The inequality is obtained from Assumption 3 that derivatives are uniformly bounded by some constant $M_1 = O(1)$. With similar arguments, we have

$$\max_k \sup_{\boldsymbol{\beta} \in \odot_k} \left| \partial S_{m,n}(\boldsymbol{\beta})/\partial \boldsymbol{\beta} - \partial S_{m,n}(\boldsymbol{\beta}_k)/\partial \boldsymbol{\beta} \right| \leq rM_2. \tag{S.51}$$

Make $r$ sufficient small such that $r(M_1 + M_2) \leq \epsilon/2$, combine (S.50), (S.51), (S.49) and (S.48), we have

$$P\left\{ \sup_{\boldsymbol{\beta} \in N_\delta(\boldsymbol{\beta}_0)} \left| \partial S_{m,n}(\boldsymbol{\beta})/\partial \boldsymbol{\beta} - \mathbb{E}\{\partial S_{i,j}(\boldsymbol{\beta})/\partial \boldsymbol{\beta}\} \right| > \epsilon \right\} < \tau. \tag{S.52}$$

The proof of Condition F2 is complete.

Condition F1 is satisfied by Assumption 3 that every term in $\partial S_{m,n}(\boldsymbol{\beta})/\partial \boldsymbol{\beta}$ is at least second-order differentiable. Since $\mathbb{E}V_{i,j}(p_0, q_0) = \mathbb{E}U_{\mathbf{Z}_{i,j}}(p_0, q_0)$, Condition F4 is satisfied. Theorem 2 in (16) can be applied. The proof is complete. ∎

## S.5 Proof of Theorem 4.4

Our proof strategy is first applying Taylor expansion to get expression $\hat{\boldsymbol{\beta}} - \boldsymbol{\beta}_0$ by $S_{m,n}(\boldsymbol{\beta}) - S_{m,n}(\boldsymbol{\beta}_0)$, then utilizing a sum to approximate $S_{m,n}(\boldsymbol{\beta}) - S_{m,n}(\boldsymbol{\beta}_0)$, finally prove the limiting distribution of the sum by triangular array central limit theorem.

By Taylor expansion, we get

$$S_{m,n}(\hat{\boldsymbol{\beta}}) - S_{m,n}(\boldsymbol{\beta}_0) \approx (\hat{\boldsymbol{\beta}} - \boldsymbol{\beta}_0) \frac{\partial S_{m,n}(\boldsymbol{\beta}_0)}{\partial \boldsymbol{\beta}},$$

Hence,

$$\hat{\boldsymbol{\beta}} - \boldsymbol{\beta}_0 \approx \left( \frac{\partial S_{m,n}(\boldsymbol{\beta}_0)}{\partial \boldsymbol{\beta}} \right)^{-1} \left( S_{m,n}(\hat{\boldsymbol{\beta}}) - S_{m,n}(\boldsymbol{\beta}_0) \right).$$

Note that

$$\mathbb{E}\{V_{i,j}(p_0, q_0) | X_i = x_i, \mathbf{Z}_j^{\bar{d}}\} = G_{\mathbf{Z}_j^{\bar{d}}}(x_i),$$

where $G_{\mathbf{Z}_j^{\bar{d}}}(\cdot)$ is the cumulative distribution function of $Y$ conditioned on $\mathbf{Z}_j^{\bar{d}}$. Thus,

$$\mathbb{E}\{G_{\mathbf{Z}_j^{\bar{d}}}(X_i)|\mathbf{Z}_i^d\} = U_{\mathbf{Z}_i^d,\mathbf{Z}_j^{\bar{d}}}(p_0, q_0). \tag{S.53}$$

Similarly, we have

$$\mathbb{E}\{V_{i,j}(p_0, q_0)|Y_j = y_j, \mathbf{Z}_i^d\} = S_{F,\mathbf{Z}_i^d}(y_j),$$

and

$$\mathbb{E}\{S_{F,\mathbf{Z}_i^d}(Y_j)|\mathbf{Z}_j^{\bar{d}}\} = U_{\mathbf{Z}_i^d,\mathbf{Z}_j^{\bar{d}}}(p_0, q_0), \tag{S.54}$$

where $S_{F,\mathbf{Z}_i^d}(\cdot)$ is the survival function of $X$ conditioned on $\mathbf{Z}_i^d$. Then, we get the sum that approximates $S_{m,n}(\boldsymbol{\beta})$,

$$S(\boldsymbol{\beta}) = \frac{1}{mn}\sum_{i=1}^{m}\sum_{j=1}^{n}\boldsymbol{\omega}_{i,j}\left[\left(G_{\mathbf{Z}_j^{\bar{d}}}(X_i) - U_{\mathbf{Z}_i^d,\mathbf{Z}_j^{\bar{d}}}(p_0, q_0)\right) + \left(S_{F,\mathbf{Z}_i^d}(Y_j) - U_{\mathbf{Z}_i^d,\mathbf{Z}_j^{\bar{d}}}(p_0, q_0)\right)\right],$$

where

$$\boldsymbol{\omega}_{i,j} = \frac{\partial U_{\mathbf{Z}_i^d,\mathbf{Z}_j^{\bar{d}}}(p_0, q_0)}{\partial \boldsymbol{\beta}}\left(U_{\mathbf{Z}_i^d,\mathbf{Z}_j^{\bar{d}}}(p_0, q_0)\bigl(1 - U_{\mathbf{Z}_i^d,\mathbf{Z}_j^{\bar{d}}}(p_0, q_0)\bigr)\right)^{-1}.$$

Combine (S.53) and (S.54), we have

$$S_{m,n}(\boldsymbol{\beta}) - S(\boldsymbol{\beta}) \to 0,$$

in probability, as $m, n \to \infty$. Applying central limit theorem for triangular arrays, we have

$$S(\hat{\boldsymbol{\beta}}) - S(\boldsymbol{\beta}_0) \to N(0, \tilde{\boldsymbol{\Sigma}}),$$

in distribution, where

$$\tilde{\boldsymbol{\Sigma}} = \lim_{m,n\to\infty}\left[\frac{1}{m^2}\sum_{i=1}^{m}\frac{1}{n^2}\sum_{j=1}^{n}\sum_{l=1}^{n}\boldsymbol{\omega}_{i,j}\boldsymbol{\omega}_{i,l}^T\mathrm{Cov}\left(G_{\mathbf{Z}_j^{\bar{d}}}(X_i), G_{\mathbf{Z}_l^d}(X_i)\right)\right]$$

$$+ \lim_{m,n\to\infty}\left[\frac{1}{n^2}\sum_{j=1}^{n}\frac{1}{m^2}\sum_{i=1}^{m}\sum_{k=1}^{m}\boldsymbol{\omega}_{i,j}\boldsymbol{\omega}_{k,j}^T\mathrm{Cov}\left(S_{F,\mathbf{Z}_i^d}(Y_j), S_{F,\mathbf{Z}_k^d}(Y_j)\right)\right]$$

Together with (S.52), the proof is complete.

## S.6  Additional Simulation

In this section, we present addition simulation results of Section 5 in the main paper.

*Case 4: Bootstrap Consistency of the Difference Estimator.* We study the coverage probability of 95% confidence interval (4.2) to support the asymptotic normality in Theorem 4.2. Let bootstrap repetition B be 1000. Samples size $(m, n)$ are chosen as: $(80, 80)$, $(150, 150)$ and $(200, 200)$. FPR ($\leq p_0$) and TPR ($\geq q_0$) constraints $(p_0, q_0)$ are $(0.7, 0.5)$, $(0.8, 0.6)$, and $(0.9, 0.7)$. The diseased subjects are generated from $N(\boldsymbol{\mu}_1, \boldsymbol{\Sigma})$ with

$$\boldsymbol{\mu}_1 = (1, 2)^\top, \text{ and } \boldsymbol{\Sigma} = \begin{pmatrix} 1 & 0.8 \\ 0.8 & 1 \end{pmatrix}.$$

The non-diseased are obtained from $N(\boldsymbol{\mu}_2, \boldsymbol{\Sigma})$ with $\boldsymbol{\mu}_2 = (0, 0)^\top$. Simulation in each setting repeats 1000 times. As shown in Table 6, coverage probabilities are around 95%, which supports the asymptotic normality in Theorem 4.2.

Table 6: Coverage probability (CP) of 95% confidence interval (4.2) for the difference estimator $\hat{\theta}(p_0, q_0)$ of two-way (TW) pAUC.

| $p_0$ | $q_0$ | $m$ | $n$ | TW pAUC CP |
|---|---|---|---|---|
| 0.7 | 0.5 | 50  | 50  | 0.948 |
| 0.7 | 0.5 | 100 | 100 | 0.952 |
| 0.7 | 0.5 | 200 | 200 | 0.950 |
| 0.8 | 0.6 | 50  | 50  | 0.952 |
| 0.8 | 0.6 | 100 | 100 | 0.950 |
| 0.8 | 0.6 | 200 | 200 | 0.950 |
| 0.9 | 0.7 | 50  | 50  | 0.949 |
| 0.9 | 0.7 | 100 | 100 | 0.948 |
| 0.9 | 0.7 | 200 | 200 | 0.951 |

Note: The region of interest is determined by FPR $\leq p_0$ and TPR $\geq q_0$. Coverage probabilities being close to 95% indicates that the asymptotic normality in Theorem 4.2 holds.

Table 7: Coverage probability (CP) of 95% confidence interval (4.1) for $\hat{U}(p_0, q_0)$ with $p_0 = 0.8, q_0 = 0.2$ in data set A, B, and C, respectively.

| $m$ | $n$ | CP A | CP B | CP C |
|---|---|---|---|---|
| 30  | 30  | 0.919 | 0.913 | 0.923 |
| 50  | 50  | 0.928 | 0.929 | 0.932 |
| 80  | 80  | 0.935 | 0.933 | 0.937 |
| 100 | 100 | 0.937 | 0.940 | 0.946 |
| 150 | 150 | 0.942 | 0.937 | 0.945 |
| 200 | 200 | 0.944 | 0.946 | 0.947 |
| 150 | 100 | 0.957 | 0.958 | 0.954 |
| 200 | 150 | 0.955 | 0.953 | 0.951 |

Note: The region of interest is determined by FPR $\leq p_0$ and TPR $\geq q_0$. The region of interest is determined by FPR $\leq p_0$ and TPR $\geq q_0$. CP being closer to 95% suggests that the asymptotic normality of $\hat{U}(p_0, q_0)$ in Theorem 4.1 holds.

*Case 5: Effect of Size on Asymptotic Normality of Estimators.* We study the effect of the restricted region's size on the coverage probability of confidence interval (4.1). FPR ($\leq p_0$) and TPR ($\geq q_0$) constraints are $(p_0, q_0) = (0.8, 0.2)$. The rest setup exactly follows Case 1 in Section 5. Combining Table 1 and Table 7, it suggests that larger size of the restricted region ensures higher coverage probability.